\documentclass[twocolumn]{aastex63}
\usepackage{multirow}
\usepackage{amsmath}
\usepackage{amssymb}

\usepackage{tikz}

\usetikzlibrary{shapes.geometric, arrows.meta, positioning}

\usepackage{graphicx}

\accepted{\today}
\usepackage{placeins}
\FloatBarrier
\submitjournal{AJ}
\usepackage{float}
\usepackage[ansinew]{inputenc}
\usepackage[american]{babel}
\usepackage{csquotes}
\usepackage{ifthen}
\usepackage{url}
\usepackage{euscript}
\usepackage{multirow}
\usepackage{nameref}
\usepackage{xcolor}
\usepackage{varioref}
\usepackage{hyperref}
\usepackage{cleveref}

\tikzstyle{process} = [rectangle, minimum width=4cm, minimum height=1cm, text centered, draw=black, fill=white!30]
\tikzstyle{decision} = [diamond, aspect=2, minimum width=3cm, minimum height=1cm, text centered, draw=black, fill=green!30]
\tikzstyle{arrow} = [thick,->,>=stealth]

\shorttitle{The 10 pc Neighborhood of Habitable Zone Exoplanetary Systems}

\shortauthors{Tisyagupta Pyne et al.}

\graphicspath{{./}{figures/}}

\usepackage{float}
\usepackage{natbib}

\begin{document}

\title{The 10~pc Neighborhood of Habitable Zone Exoplanetary Systems: \\
Threat Assessment from Stellar Encounters \& Supernovae}

\correspondingauthor{Ravinder K Banyal}
\email{banyal@iiap.res.in}

\author {Tisyagupta Pyne}
\affiliation{Indian Institute of Astrophysics, Koramangala 2nd Block, Bangalore 560034, India}
\affiliation{Integrated Science Education and Research Centre, Visva-Bharati University, Santiniketan, 731235, India}

\author[0000-0003-0799-969X]{Ravinder K. Banyal}
\affiliation{Indian Institute of Astrophysics, Koramangala 2nd Block, Bangalore 560034, India}

\author[0000-0003-1371-8890]{C. Swastik}
\affiliation{Dipartimento di Fisica, Università degli Studi di Milano, Via Celoria 16, 20133 Milano, Italy}
\affiliation{Indian Institute of Astrophysics, Koramangala 2nd Block, Bangalore 560034, India}
\affiliation{Pondicherry University, R.V. Nagar, Kalapet, 605014, Puducherry, India}

\author[0000-0001-8845-184X]{Ayanabha De}
\affiliation{Indian Institute of Astrophysics, Koramangala 2nd Block, Bangalore 560034, India}


\begin{abstract}
The habitability of a planet is influenced by both its parent star and the properties of its local stellar neighborhood. Potential threats to habitability from the local stellar environment mainly arise from two factors: cataclysmic events such as powerful stellar explosions and orbital perturbations induced by close stellar encounters. Among the 4,500+ exoplanet-hosting stars, about 140+ are known to host planets in their habitable zones. In this study, we use \textit{Gaia DR3} data to investigate the 10~pc stellar neighborhood of the 84 habitable zone systems (HZS) closest to the Sun. We assess the possible risks that local stellar environment of these HZS pose to their habitability.  In particular, we find that HD~165155 has a high stellar density around it, making it likely to experience at least one flyby encounter within a span of 5~Gyr. We also identified two high-mass stars ($M \geq 8 M_\odot$) as potential progenitors of supernovae, which could threaten the long-term survivability of habitable zone systems HD~48265 and TOI-1227. Further, to quantify the similarity between habitable zone stars and the Sun, as well as their respective 10~pc stellar environments, we employ various astrophysical parameters to define a Solar Similarity Index (SSI) and a Neighborhood Similarity Index (NSI).  Our analysis suggests that HD~40307 exhibits the closest resemblance to the solar system, while HD~165155 shows the least resemblance. 
\end{abstract}.


\section{Introduction}\label{sec1}
Finding a habitable world is one of the primary goals of exoplanetary research. The study of habitability is a rapidly growing field in exoplanet science, with over 150 confirmed discoveries of planets residing in the habitable zones' of stars already made \citep{Fujii_2018,doi:10.1089/ast.2017.1729,Glaser_2020,Lisse_2020,Hill_2023}. Traditionally, the habitable zone (HZ) is defined as the annular region around a star where liquid water can exist on a planet under sufficient atmospheric pressure.

In modern lexicon, habitable zones are typically categorized into two broad types \citep{KASTING1993108,under2003}. The first is the conservative habitable zone, which is defined by an inner boundary where the intense radiant energy from the star may induce a runaway greenhouse effect, resulting in the vaporization of surface water. Its outer boundary is determined by the distance from the central star at which a planet's cloud-free CO$_2$ atmosphere can maintain a surface temperature of 273~K. In contrast, the second type, known as the optimistic habitable zone (OHZ), encompasses regions receiving radiation levels ranging between those experienced by Mars $\sim$4 billion years ago and Venus around 1 billion year ago. \citep{Kopparapu_2013,Kopparapu_2014, Ware_2022}. 

In the galactic context, the conditions favourable for life are also dependent on spatial and temporal location of star-planet systems within the Milky Way \citep{gon2001,lineweaver2004galactic}. Aspects of galactic habitability include the radiation threat from high-energy events like Supernovae (SNe), Gamma-ray bursts (GRBs), presence of heavy elements that are crucial for the formation of rocky planets, star-forming regions and epoch of planet formation in the galaxy \citep{spitoni2017galactic, spinelli2021best, swa2023, Spinelli_2023,2024AJ....167..270S}.  

While the concept of HZ is vital in the search for habitable worlds, the stellar environment of the planet also plays an important role in determining longevity and maintenance of habitability. In particular, a planet's habitability can be greatly influenced by the type and distribution of stars surrounding the exoplanet hosting star. Studies have shown that a high rate of catastrophic events, such as supernovae and close stellar encounters in regions of high stellar density, is not conducive to the evolution of complex life forms and the maintenance of habitability over long periods \citep{Torres_2013,spitoni2017galactic,spinelli2021best,Spinelli_2023}. Important as they are, these theoretical ideas have not been tested against the observed ensemble of extrasolar planets. 

The growing number of exoplanet discoveries has enabled researchers to gather robust statistics, warranting further investigation into their stellar environments \citep{nar18, 2021AJ....161..114S, uni22, nar23,ban24}. The latest census of confirmed exoplanetary systems and their astrophysical properties are available at various public archives \citep{Schneider_2011,Akeson_2013,Han_2014,nea12}. Out of the 5500+ discovered exoplanets so far, 146 stars are known to host 158 rocky and gaseous planets within their HZ, as documented in the Catalog of Habitable Zone Exoplanets \citep{2012PASP..124..323K,Hill_2023}. Some of these systems are potential targets for detailed atmospheric characterization and the detection of bio-signatures in current and future missions \citep{sta2014,tin2018,red2024}. 

In this work, we focus on the local stellar environment of stars hosting habitable zone exoplanets. More specifically, we use \textit{Gaia DR3} archive to analyze the 10~pc neighborhood (stars within a sphere of 10~pc radius) surrounding the known habitable zone systems (HZS). This involves extracting sources from the \textit{Gaia DR3} archive with measured parallaxes from $\varpi \approx 750~\textrm{mas}$ to $\varpi \approx 4.5~\textrm{mas}$ which correspond to the nearest (Proxima Centauri) and the farthest (Kepler-296) HZS. The rationale for selecting a 10~pc region is based on studies indicating that if a star within this range evolves into a Type-Ia/Ibc SNe, it would produce X-rays and $\gamma$-rays with sufficient fluence (total energy emitted by a SNe per unit area) to significantly disrupt a planet's ozone layer \citep{Spinelli_2023}. Additionally, the 10~pc neighborhood of the solar system, well-studied using {\it Gaia DR3} by \cite{reyle202110}, provides a valuable reference for exoplanet demographic comparison.
We further explore the likelihood of stellar encounters and supernovae that could pose a threat to the habitability of these systems. We also compare the 10~pc neighborhood of these HZS with the 10~pc solar neighborhood sample compiled by \cite{reyle202110} and \cite{Gaia_10pc_Update}. 

We define the Solar Similarity Index (SSI) as a metric for comparing the properties of the habitable zone planet-hosting stars with those of the Sun. Similarly, the Neighborhood Similarity Index (NSI) is a metric used to compare the properties and distribution of objects within their respective volumes of 10~pc radius.  These indices are then used to determine which habitable systems have the closest resemblance to our solar system and its 10~pc environment. The closer the index is to 1, the more similar it is. The stellar and planetary data (see Table~\ref{tab:1}) for this study are obtained from the \textit{Gaia DR3} and the NASA Exoplanet Archive (NEA) \citep{gai2023,Akeson_2013,nea12}.

From here on, the paper is organized as follows: In Section~\ref{sec2}, we describe our primary sample of HZS and the construction of their 10~pc neighborhoods. In Section~\ref{sec3}, we discuss the influence of the stellar neighborhoods on habitability and present our main results. In Section \ref{exhaust}, we address the incompleteness of {\it Gaia} data and its implications for our findings. Finally, in Section~\ref{sec5} we summarize and conclude this work.

\section{Sample Selection}\label{sec2}
In this section, we describe the process of our sample selection which mainly consists of two parts. The first part involves selecting 146 planet-hosting stars from the Habitable Zone Gallery\footnote{\url{https://hzgallery.org/}}, which constitutes our \textit{primary sample} of HZS. For the second part, we used the \textit{Gaia DR3} catalog to select stars within the 10~pc neighborhood of 144 HZ host stars\footnote{Parallax measurements were not available for two of the systems, Kepler-1652 and Kepler-1410.}. The subset of stars in each 10~pc HZ neighborhood  was curated from a larger 25~pc dataset, as explained in Section \ref{Bootstrap}.

\subsection{Habitable Zone Gallery}\label{HZG}

The Habitable Zone Gallery is an online catalog of known exoplanets and their orbital parameters \citep{2012PASP..124..323K,Hill_2023}, constructed using information from larger databases such as Exoplanet Data Explorer\footnote{\url{http://exoplanets.org/table}} and the NEA\footnote{\url{https://exoplanetarchive.ipac.caltech.edu/}}.  The Gallery demarcates the habitable zones of each planet-hosting star as defined by \cite{Kopparapu_2013,Kopparapu_2014}.  It also calculates the fraction of time a planet spends in the habitable zone, which can vary from 0\% to 100\% depending on the planet's eccentricity and orbital distance. This catalog is regularly updated and it serves as a valuable resource for researchers studying the habitability of exoplanets. 

\begin{figure}[!t]
\centering
\includegraphics[width=1 \columnwidth]{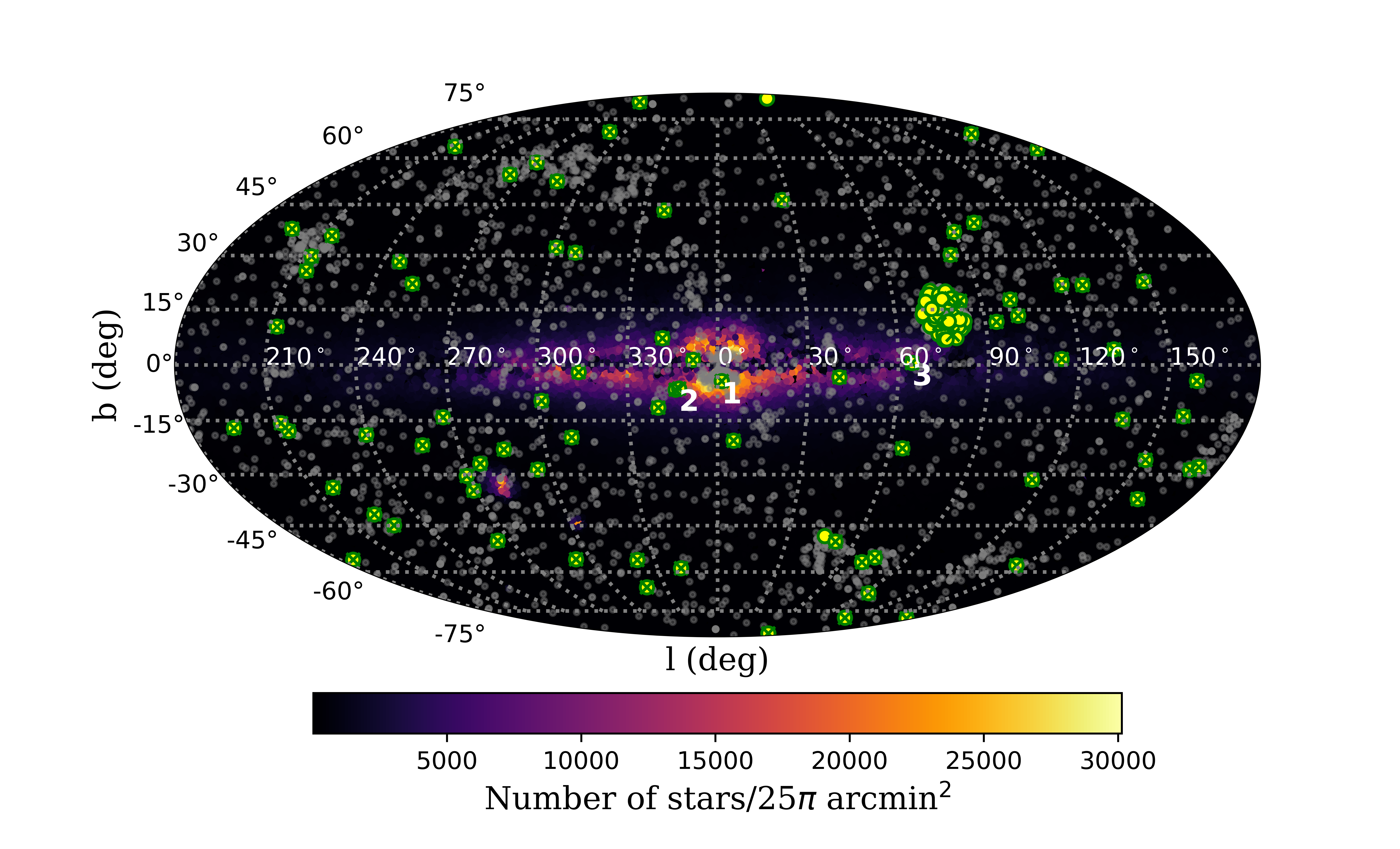}
\caption{Sky positions of exoplanet-hosting stars projected on Molleweide map. HZS are denoted by yellow-green circles, while the remaining population of exoplanets is represented by gray circles. The studied sample of 84 HZS, located within 220 pc of the Sun, is represented by crossed yellow-green circles. The three high-density HZS located near the galactic plane are labeled 1, 2 and 3 in white. The colorbar represents the stellar density, i.e., the number of stars having $G\geq15$ within a radius of 5~arcmin.}
\label{POS}
\end{figure}

Our primary sample of HZS, drawn from the Habitable Zone Gallery, comprises 146 systems hosting 158 planets whose orbits fully reside within the optimistic habitable zone of their host stars. Of these 158 planets, 35 are presumably rocky planets ($\leq$2R$_\oplus$), 122 are gaseous planets, and 1 has an undetermined mass and radius. While the giant planets themselves are inhospitable, they may host rocky exomoons orbiting them under favorable conditions for life \citep{Heller_2012,Heller_Barnes_2013}. Additionally, the moons of giant planets located at the outer edge of the habitable zone could generate sufficient energy through tidal heating \citep{Heller_Barnes_2013,Heller_Armstrong_2014,Hill_2018,Hill_2023}. The astrophysical parameters of these systems were obtained from the NASA Exoplanet Archive. Figure~\ref{POS} shows the all-sky distribution of exoplanet hosting stars in a Mollweide projection. The nearest HZS to the Sun is Proxima Centauri, located at a distance of 1.3~pc, while the farthest is Kepler-1636, at a distance of 2.2~kpc.

\subsection{Curating the 10~pc neighborhood}\label{Bootstrap}

\begin{figure}[t!]
\centering
\includegraphics[width=1.\columnwidth]{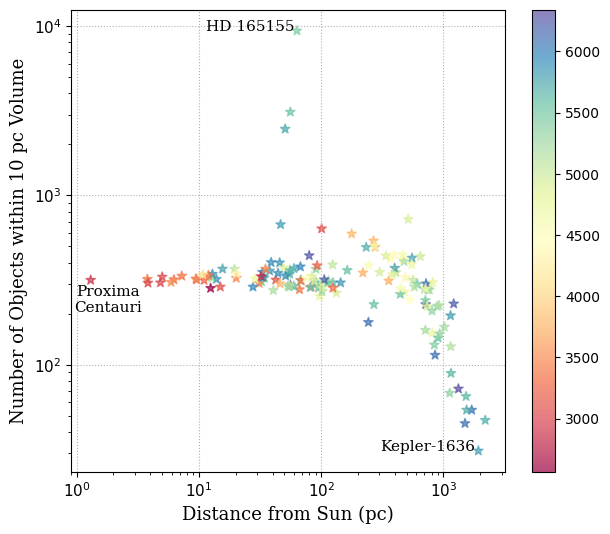}
\caption{The 10~pc neighborhood star count for 144~HZS and their distance from the Sun. The colorbar represents the effective temperature $T_{\textrm{eff}}$ (K) of HZ stars.}
\label{obno}
\end{figure}

For studying the 10~pc environment of HZS using \textit{Gaia DR3}, we have to contend with two major issues: 
\begin{itemize}
    \item The ambiguity of stars belonging to the 10~pc region due to distance and magnitude dependent parallax errors (see Appendix~\ref{derive}), and 
    \item The incompleteness of \textit{Gaia} data, i.e., \textit{Gaia}'s inability to detect sources outside $21\lesssim G \lesssim 3$ magnitude range and additional cuts on various astrophysical parameters, impacting its overall dataset \citep{reyle202110, gai2023}. 
\end{itemize}
The relative parallax errors in \textit{Gaia} data can significantly affect distance estimates and introduce ambiguity in defining the 10~pc boundary. This means that a strict 10~pc neighborhood sample returned by a \textit{Gaia} query (see Figure~\ref{obno}) may include some stars with large parallax errors that could, in reality, be outside  the 10 pc sphere, or it may exclude stars that are truly within the 10~pc boundary. Therefore, to address this issue, the 10~pc stellar environment for each HZ system was constructed from a superset of neighboring stars distributed within a sphere of radius 25~pc. For this, we used a simple bootstrapping approach that alleviates the need to define an exact 10~pc boundary and helps constrain the uncertainties of astrophysical parameters of neighborhood stars at the ensemble level, which we later use for quantifying the similarity indices. 

\begin{figure}[t!]
\centering
\includegraphics[width=1.\columnwidth]{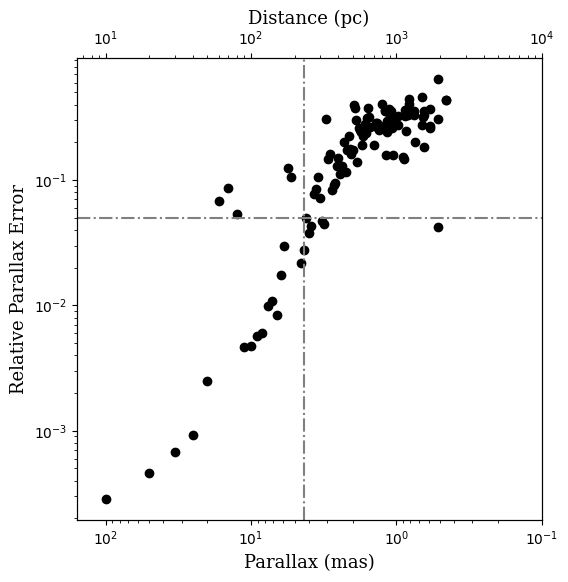}
\caption{Relative parallax errors of Gaia detected stars binned at 10~pc intervals as a function of increasing distance from the solar system.} 
\label{extents}
\end{figure}

Apart from the stated ambiguity in the star count within the 10~pc sample, Figure \ref{obno} shows a decline in the number of neighborhood stars around HZS beyond a distance of $\sim$~300~pc. This selection bias (under count) arises due to the incompleteness of {\it Gaia} data, which is discussed later in Section \ref{exhaust}. Figure~\ref{extents} shows how the relative parallax error of \textit{Gaia} detected stars increases with decreasing parallax (i.e., with increasing distance from Sun). Since our goal is to study the 10~pc region around HZ stars, if the distance error were to exceed 10~pc, defining the neighborhood would become arbitrary. For the \textit{Gaia} detected stars shown in Figure~\ref{extents}, the relative parallax error exceeds 5\% beyond 220~pc ($\varpi \approx 4.5$~mas) which corresponds to a distance error $\sim$10~pc. Hence, we only consider 84~HZS from our primary sample of 146~HZS, which are within a distance of 220~pc from the Sun. 
By limiting our sample to 220~pc, we also ensure a nearly complete detection of objects within their 10~pc neighborhood, including the cool-dwarfs down to spectral type M6, by {\it Gaia} (see Section \ref{exhaust}). 

\begin{figure}[t!]
    \centering
    \includegraphics[width=1 \columnwidth]{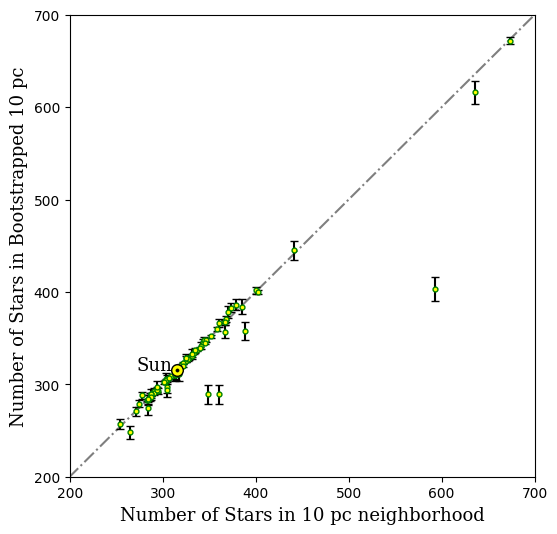}
    \caption{Comparison of number of stars returned by Gaia ADQL Query. (x-axis) within 10~pc neighborhoods and the 10~pc mean count of stars (y-axis) obtained after $10^5$ bootstrap runs on a larger dataset of stars within 25~pc region. The errorbars represent $\pm1\sigma$ uncertainly in  the count of stars after bootstrapping. For brevity, the plot is curtailed at 700 star count, excluding three sources with more than 2000 stars.}
    \label{before_after}
\end{figure}

To curate a 10~pc sample, we begin by issuing a standard ADQL query on the {\it Gaia DR3} catalog to select stars within 25~pc of a HZS. The query returns all stars around a HZS with distances $\leq$~25~pc and their astrophysical parameters (see Appendix~\ref{10pc_code}). Next, for each HZS, we generate 100,000 random realizations of 10~pc neighborhood stars from this superset by using the bootstrapping method described in Appendix~\ref{derive}. The astrophysical parameters\footnote{See Table \ref{tab:1} for the astrophysical parameters obtained from the statistical inferences of sampling distributions.} and their associated uncertainties are inferred from the sampling distribution of the bootstrap. 
In Figure~\ref{before_after}, we compare the average star count obtained from the bootstrap method and the star count obtained directly from {\it Gaia}'s 10~pc query for 84~HZS. Apart from a few outliers, the differences are not significant. However, the bootstrap method provides more reliable estimates of the astrophysical parameters of neighborhood stars.

Finally, the total star count within the 10~pc region surrounding the 84 HZS is found to be $\sim$~36,000 stars. This dataset does not include the brighter stars due to the lower magnitude limit (G$\sim$3) of Gaia. Finding  the bright stars in the vicinity of HZS is crucial for assessing the threat to habitability from supernovae. Therefore, to further complement the 10~pc dataset we separately searched the Hipparcos\footnote{\href{https://heasarc.gsfc.nasa.gov/W3Browse/all/hipnewcat.html}{Hipparcos Catalog 2007}} Catalog \citep{Hipparcos1997,Hipparcos_2007}, for bright stars. We found 34 bright stars with \verb|hip_mag|~$<$3 belonging to the 10~pc neighborhood of different HZS. Only two of these 34 bright stars, $\alpha$-Carinae (around HD 48265) and $\alpha$-Muscae (around TOI-1227), are massive enough ($\geq$~8~M$_\odot$) to pose a noteworthy threat to the habitability (see Section \ref{sne}).

\section{Results and Discussions}\label{sec3}
While numerous risks, such as activity-induced stellar winds and superflares from a host star can compromise a planet's habitability \citep{atmosphere_loss,magnet_plasma,magnetospheres}, here, we examine the possibility of any significant impact from stellar encounters and supernova explosions in the surrounding stellar environment of HZS. These events can significantly alter the habitability of planets by displacing them out of their HZ or by disrupting  their atmosphere. Studies have shown that the frequency and proximity of such events has a critical role in the habitability of exoplanets within our galaxy \citep{lineweaver2004galactic,Torres_2013,spitoni2017galactic,10.1093/mnras/stz1794,10.1093/mnras/staa1622,spinelli2021best,10.1093/mnras/stac3705,Spinelli_2023}. These factors further determine  the long-term viability of habitable conditions on an exoplanet.

\subsection{Stellar Encounters}\label{fly-bymeth}
Stellar encounters can impact exoplanetary systems in various ways \citep{hor20,dav22}. A passing star's gravitational influence can perturb distant objects, such as those in the Oort Cloud, pushing them into highly elliptical orbits that may lead to collisions with inner planets. Closer encounter with neighboring stars can directly destabilize planetary orbits, potentially causing planets to migrate inward, outward, or even escape the system entirely. Such disruptions can also alter the eccentricity and inclination of planetary orbits, reducing the time planets spend in the habitable zone (HZ) and threatening their habitability \citep{Wang_2020,rickman_2023}. In some cases, these perturbations may trigger mechanisms like the Kozai-Lidov effect, leading to oscillations in eccentricity, angular momentum exchange, and changes in orbital inclinations \citep{kozai-lidov,Cai_2017}. These orbital shifts may result in variations in stellar insolation flux, which, in turn, can affect a planet's atmosphere, climate, and potential habitability.  Additionally, the destabilization of an outer planet could trigger a cascading effect inward, amplifying instability across planets in the inner orbits. The overall impact of these encounters depends on the mass and proximity of the passing star, as well as the architecture of the planetary system.

\cite{Torres_2013} examined the effect of encounters on habitability in various stellar environments of the Milky Way. They simulate different regions by estimating stellar densities and dispersion velocities, creating a model to approximate the number of close flyby events that can potentially alter the orbital dynamics. 
They showed that a 1~M$_\odot$ star passing at a distance of 200~AU can perturb another stellar system with a radius of 100 AU, which has an Oort-like cloud surrounding it.
This model is based on the neighborhood's stellar density, dispersion velocity and evolution time. According to \cite{Torres_2013}, the number of encounters, $N_e$ is given by the following equation,
\begin{equation}\label{s_enc}
    N_e = 4\pi nvT_eR^2
\end{equation}
where $N_e$ is the number of encounters, $R$ is the 
radius of the stellar system ($b = 2R$ is the impact parameter\footnote{The distance of closest approach made by the flyby star to the host star, denoted by $b$.}), $n$ is the stellar density in pc$^{-3}$, $v$ is the dispersion velocity of the system and $T_e$ is the evolution time of the system. Although, the evidence for Oort-like clouds around exoplanetary systems is still lacking, it is not uncommon to speculate about their existence \citep{Torres_2013,Oort_Baxter,Oort_Cloud_Zwart}. The possibility of planetary bodies residing in Oort-like regions, which can extend up to 100~AU from the host star, has also been suggested \citep{Oort_Exoplanets}. 

We apply the formalism of Equation~\eqref{s_enc} to real data (i.e., our HZS sample) and estimate the frequency of gravitational encounters from stars in the 10~pc region of HZS. We used the proper motions and radial velocity of neighborhood stars to calculate the $U$, $V$ and $W$ components using PyAstronomy\footnote{LSR= (8.5, 13.38, 6.49) \citep{UVW_LSR}} 
\citep{pya,swastik2023age}. $U$ denotes the velocity toward the galactic center, $V$ denotes the velocity in the direction of galactic rotation and $W$ denotes the velocity  toward the north galactic pole. Then, we find the dispersion velocity $v$ from: $v = \sqrt{\sigma_U^2 + \sigma_V^2 + \sigma_W^2}$, where $\sigma$ is the dispersion of the respective components.

\begin{figure}[!t]
\centering
\includegraphics[width=1.\columnwidth]{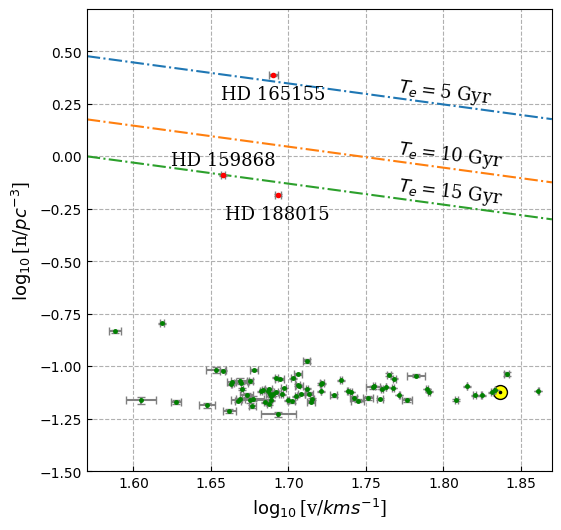}
\caption{A log-log diagram of stellar density versus velocity dispersion. Slanted lines indicate the locus of a single encounter ($N_e = 1$) in stellar density-dispersion parameter space for time scales of 5, 10 and 15 Gyr and $b = 150$~AU. 
Three high stellar density environment ($>$~0.4~pc$^{-3}$) are shown by red symbols while the remaining low-encounter HZ systems are indicated by green symbols.}
\label{encounter}
\end{figure}

In these calculations, we used the median mass\footnote{Masses have been estimated from the updated (2022) table for spectral sequence based on \href{https://www.pas.rochester.edu/~emamajek/EEM_dwarf_UBVIJHK_colors_Teff.txt}{Pecaut \& Mamajaek 2013}}  of neighborhood stars to be 0.3~M$_\odot$, leading to an impact parameter of $b=150$~AU. This is inline with peak of the stellar mass function (0.3--0.4~M$_\odot$), which is consistent with the general prevalence of low-mass M-type stars in the Milky Way and the solar neighborhood \citep{reyle202110}. The impact parameter $b=150$~AU is determined by calculating the distance at which a neighboring star with a mass of 0.3~M$_\odot$  exerts a gravitational force on the exoplanetary Oort cloud that is equal to the force exerted by a 1~M$_\odot$  star at a distance of $b=200$~AU. Figure~\ref{encounter} shows a log-scaled plot of stellar density versus dispersion velocity for our sample of 84~HZS. The slanted lines in Figure~\ref{encounter} denote the single encounter thresholds for time scales of 5, 10 and 15~Gyr. This means that for any HZS to experience at least one encounter within a certain time frame, it must lie on or above the slanted line. Since the stellar density around most HZS shown in Figure~\ref{encounter} is low ($< 0.2 $~pc$^{-3}$) they are positioned well below the 15~Gyr line and  face a negligible threat to habitability from stellar encounters.

Among the 84 systems, three habitable zone stars, namely HD~165155, HD~159868, and HD~188015, are residing in a region with unusually high stellar density ($>$~0.6~pc$^{-3}$) environment compared to Sun and other HZS. This is not surprising, given that these HZS are located near the galactic plane (see Figure~\ref{POS}). However, to rule out erroneous star counts from background contamination and other spurious sources, a further assessment of these three systems is provided in Appendix~\ref{high-dense-nbh}.

Notably, HD~165155 is a G8V star with the planet HD~165155~b in the habitable zone \citep{HD_165155}. This system has the highest stellar density ($n \approx ~2.45$~pc$^{-3}$) in our dataset, with over $10235\pm67$ stars  within a 10~pc radius and velocity dispersion $\approx$49~kms$^{-1}$.  Given its high-density environment, HD~165155 is expected to undergo at least one stellar encounter within  5~Gyr. In contrast, the other two systems, HD~188015 and HD~159868, have a small likelihood ($N_e<1$) of experiencing stellar encounters due to their lower stellar density.  

Studies on stellar encounters have discussed the evolution of planetary systems in highly dense environments by simulating scenarios with 2000, 8000 and 32000 stars within a virial radius\footnote{The radius within which objects exist in a gravitationally bound state.} of 1~pc over a span of 50~Myr \citep{Cai_2017}. They discuss the survival rates of planets in such environments indicating a clear correlation between survival rates and decreases in stellar density. The work of  \cite{Arbab_2021} highlighted that the mass and velocity of the flyby stars are crucial in determining the encounter dynamics. Some studies also discuss the formation of HZ planets in clustered environments and their lifetimes with respect to the stellar mass and the stellar densities of their neighborhoods \citep{Ovelar_2012}. 

For our HZS sample, the encounter rate was calculated for different $n$ and $v$. Which means the dispersion velocity $v$ was determined for brighter stars, $G \lesssim 15$, for which radial velocity data was available from {\it Gaia DR3} (see Section~\ref{exhaust}). Since RV data were not available for the fainter stars, the dispersion velocity obtained is only the lower-limit.  However, due to the shallow slope of the single encounter lines in Figure~\ref{encounter}, the underestimation of dispersion velocity does not affect the encounter rate in any significant way.

\subsection{Assessing Threat from SNe Explosions}\label{sne}
High-energy particles and radiation arriving from distant regions of space can potentially damage the atmosphere of Earth-like planets or exomoons with an Earth-like atmosphere. Such radiation originate from high-energy transient phenomena, such as $\gamma$-ray bursts and SNe, which involve a brief period of intense radiation that diminishes over time. Our primary focus is to investigate the effects of SNe on the atmospheres of exoplanets or exomoons assuming their atmospheres to be Earth-like \citep{Thomas_Melott_2006,Melott_2011,Perkins_2024}. The fluence $F$ received by a planet from a high-energy transient event is given by,
\begin{equation}\label{eq:9}
    F = \frac{\langle E \rangle}{4\pi r^2}
\end{equation}
where $\langle E \rangle$ is the characteristic energy of the event and $r$ is the distance of SNe from the planet \citep{spinelli2021best,Spinelli_2023}.
The severity of the ozone depletion would depend on the fluence received by the planet. A fluence $\gtrsim 10\textrm{kJm}^{-2}$ would deplete the ozone layer and make the planet vulnerable to harmful radiation and might render the planet uninhabitable \citep{Thomas_2005,Thomas_2005a,Melott_2011,Horvath_Galante_2012,Spinelli_2023}. In our study, we primarily focus on SNe-Ibc and SNe-II since their occurrence is relatively higher than other sources of X and $\gamma$ radiation such as GRBs and SNe-Ia \citep{SNe_Rates1,Spinelli_2023}.

It is well established that a Type Ib,c or a Type-II SNe requires a progenitor star with a mass exceeding 8~M$_\odot$.
While SNe-Ia are formed from binary accreting systems involving a degenerate star, SNe-Ibc and SNe-II are formed from core-collapse \citep{Spinelli_2023}. The typical characteristic energy of an SNe ranges from a minimum of $10^{33}$~kJ (SNe-II) to a maximum of $10^{37}$ kJ (SNe-I) \citep{spinelli2021best,Spinelli_2023}.
Therefore, we focus on identifying stars with mass $\geq$~8~M$_\odot$ within 10~pc of each HZ system, as outlined by \cite{Spinelli_2023}. \textit{Gaia}'s Final Luminosity Age Mass Estimator (FLAME) provides an estimate of stellar mass \citep{Gaia_apsis_I,Gaia_apsis_II,Gaia_apsis_III}, but it is not available for all the stars in our sample. We use the mass-luminosity relationship described by \cite{mass_lum} in Equation~\eqref{eq:7} to estimate masses for stars in our dataset that lack FLAME-derived mass estimates.

\begin{figure}[t!]
\centering
\includegraphics[width=1.\columnwidth]{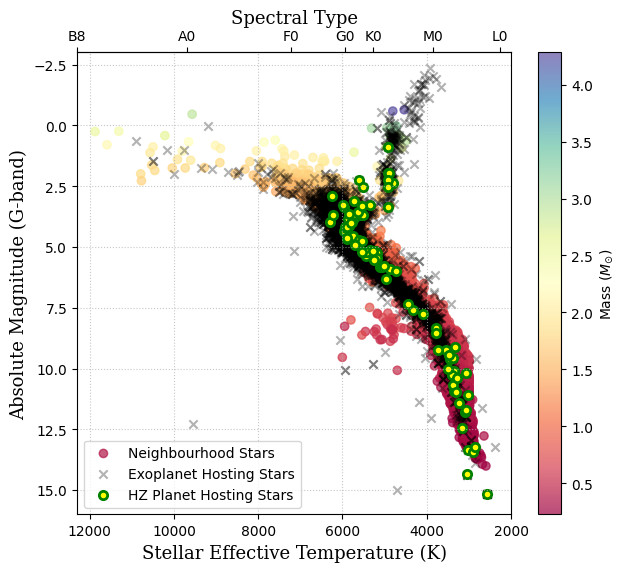}
\caption{HR diagram of all GAIA detected stars in 10~pc region of 84~HZS  and the known exoplanet hosting stars. The neighborhood stars are color-coded according to their estimated mass.}
\label{HRD}
\end{figure}

\begin{figure}[!t]
\centering
\includegraphics[width=1.\columnwidth]{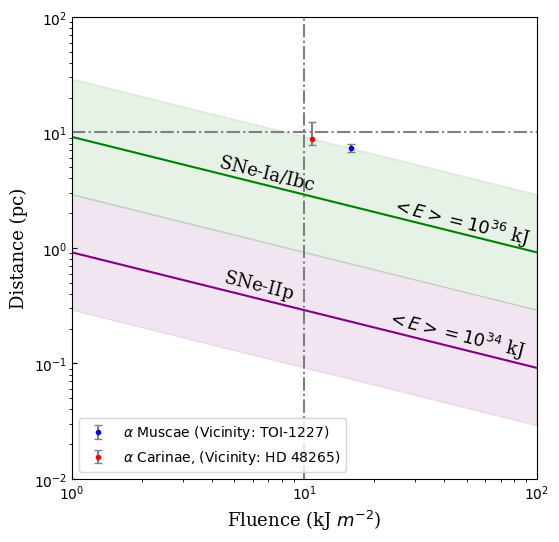}
\caption{Fluence received from  SNe explosions as a function of distance.  The purple and green bands depict the range of energy released from different SNe.  The shaded region to the right of the vertical line receives a fluence in excess of 10~kJm$^{-2}$, which is considered harmful to habitability. Locations of two high-mass stars ($M>8~M_\odot$), which are potential progenitors of supernovae, are also shown for comparison.}
\label{supernova}
\end{figure}

\begin{equation}\label{eq:7}
    \frac{M}{M_\odot} = \left(\frac{L}{L_\odot}\right)^{\frac{1}{4}} \left(\frac{1}{3}\right)^{{\left({T_{\textrm{eff}}}/{T_\odot}\right)}^{\frac{1}{3}} - 1} 
\end{equation}

Here, $L$ is the luminosity, $M$ is the mass and $T_{\textrm{eff}}$ is the surface temperature of the star. We used Pogson's formula in Equation~\eqref{eq:7} to calculate luminosity from absolute magnitudes \citep{1856MNRAS..17...12P,pogson_use_2018} for stars without FLAME luminosity.  In our analysis of Gaia DR3 dataset we do not find any star with $M\geq$6~M$_\odot$ surrounding HZS.

Figure~\ref{HRD} is the Hertzsprung-Russell (HR) diagram of our full Gaia dataset of neighborhood stars. To convert from effective temperature to spectral type, we referred to the updated spectral sequence\footnote{\href{https://www.pas.rochester.edu/~emamajek/EEM_dwarf_UBVIJHK_colors_Teff.txt}{Updated (2022) table for spectral sequence based on Pecaut \& Mamajaek 2013}} based on \cite{Pecaut_Mamajek_2013}. The HR diagram shows that the neighborhood stars in our sample range from L0 to B8-type. Since stars brighter than ($G\leq 3$) are absent in the Gaia catalog, we searched the {\it Hipparcos} Catalog\footnote{\href{https://heasarc.gsfc.nasa.gov/W3Browse/all/hipnewcat.html}{Hipparcos Catalog 2007}} \citep{Hipparcos1997,Hipparcos_2007} and found 2~HZS (TOI-1227 and HD 48265), having high-mass stars ($M>$~8 M$_\odot$), namely $\alpha$-Muscae and $\alpha$-Carinae respectively, within their 10~pc vicinity. These stars are potential SNe progenitors. We consider the evolution of these high-mass stars into SNe-Ibc as the worst case scenario from the  viewpoint  of habitability of the central HZS. The maximum characteristic energy, $\langle E \rangle$ that a SNe-Ibc would produce is $\sim 10^{37}~\textrm{kJ}$.

The variation of fluence received from SNe explosions as a function of distance and characteristic energy released  is illustrated in Figure~\ref{supernova}. In the figure we depict two high-mass stars that are located at a distance $\lesssim 10$~pc from HZS TOI-1227 and HD~48265.  A resulting stellar explosion (SNe-Ibc) releasing a maximum energy $\langle E \rangle = 10^{37}$~kJ would produce a lethal fluence~$\geq$~10~kJm$^{-2}$ of X-rays/$\gamma$-rays \citep{Spinelli_2023} within 10~pc sphere. The planets in these two systems, namely TOI-1227 b \citep{TOI-1227} and HD 48265 b \citep{HD_48265} are gaseous. Although these planets themselves cannot have habitable conditions, they might host exomoons that could have Earth-like atmospheres susceptible to these effects. For an earth-like atmosphere this could deplete  $\sim$68\% of the ozone layer \citep{Thomas_Melott_2006,spinelli2021best}. 

Note that $\gamma$-ray bursts can emit immense energy, affecting planets as far as $\sim 1$~kpc away \citep{spinelli2021best,Spinelli_2023}. However, due to the unpredictability and rarity of $\gamma$-ray burst events, we do not consider them within the scope of present work. Similarly, predicting whether a main-sequence white dwarf binary will undergo a Type Ia SNe is challenging, as it depends on factors like their proximity and orbital dynamics of a compact binary system. In contrast, the likelihood of Type Ibc and Type II supernovae is higher in stars with masses $\geq 8M_\odot$. In a worst-case scenario, if a high-mass star is stripped of its hydrogen envelope, it will evolve into a SNe-Ib. If both hydrogen and helium are depleted, it will become a SNe-Ic. However, it is rare for a high-mass star to evolve into these categories by losing sufficient hydrogen from its outer layers. In most cases, a high-mass star undergoes core collapse, producing less energetic Type-II supernovae (SNe-II), which indicates their impact is less damaging beyond 5~pc \citep{Melott_2017}. None of our HZS have high-mass stars within a 5~pc region. Recent studies also suggest that a SNe explosion occurring even at  distances up to 20~pc could be lethal to a planet's habitability \citep{SNe_Updated_2023}.

\subsection{Similarity Indices}\label{simil}

\begin{figure}[!t]
\centering
\includegraphics[width=1.\columnwidth]{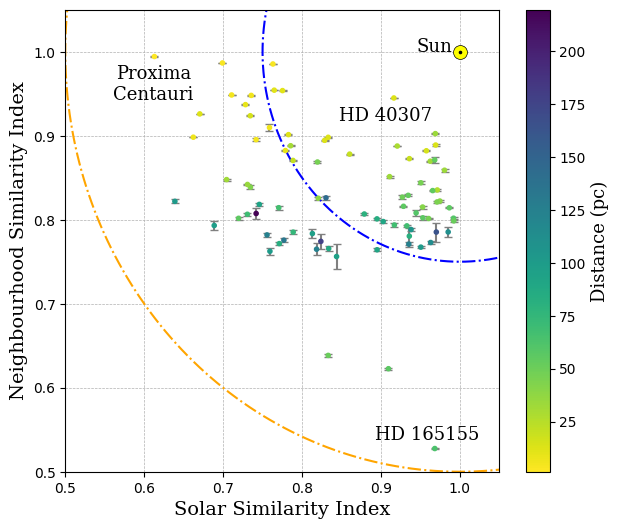}
\caption{Similarity index based comparison of habitable zone systems with solar system. The colorbar represents the distance of HZS from Sun in pc. The blue and orange semicircles centered on the Sun correspond to  0.75 and 0.50 similarity in NSI--SSI plane.}
\label{ssi-nsi}
\end{figure}

\begin{table*}
\centering
\begin{tabular}{llll}
\hline
\textbf{Column Name} & \textbf{Unit} & \textbf{Description} & \textbf{Example Value} \\
\hline
\verb|Host_Star| & - & Name of host star & HD 165155 \\
\verb|designation| & - & Gaia DR3 identifier & Gaia DR3 4050... \\
\verb|ra| & degrees & Right ascension & 271.49 \\
\verb|dec| & degrees & Declination & -29.92 \\
\verb|parallax| & mas & Parallax & 15.78 \\
\verb|pm_ra| & mas/yr & RA proper motion  & 76.89 \\
\verb|pm_dec| & mas/yr & Dec proper motion  & -1.57 \\
\verb|radial_velocity| & kms$^{-1}$ & Radial velocity  & 15.3 \\
\verb|phot_g_mean_mag| & G-mag & Photometric G-band magnitude & 9.22 \\
\verb|s_teff| & K & Surface temperature of host star from NEA & 5426 \\
\verb|s_logg| & $\log(\textrm{cms}^{-2})$ & Surface gravity of host star from NEA & 4.49 \\
\verb|s_mul| & - & Number of stars in the system & 1 \\
\verb|absolute_mag| & G-mag & Absolute magnitude in G-band & 5.21 \\
\verb|S.S.I| & - & Solar similarity index & 0.97 \\
\hline
\verb|Dispersion_Velocity_log| & $\log(\textrm{kms}^{-1})$ & Log of dispersion velocity of neighborhood stars & 1.690 \\
\verb|Dispersion_Velocity_log_std| & $\log(\textrm{kms}^{-1})$ & Uncertainty in log of dispersion velocity of neighborhood stars & 0.003 \\
\verb|Object_Den_log| & $\log(\textrm{pc}^{-3})$ & Log of object density of neighborhood stars & 0.388 \\
\verb|Object_Den_log_std| & $\log(\textrm{pc}^{-3})$ & Uncertainty in log of object density of neighborhood stars & 0.003 \\
\verb|Number| & - & Number of neighborhood stars & 10235 \\
\verb|Number_std| & - & Uncertainty in number of neighborhood stars & 67 \\
\verb|D_Teff| & K & Dispersion in T$_\textrm{eff}$ for neighborhood stars & 671 \\
\verb|M_Teff| & K & Median of T$_\textrm{eff}$ for neighborhood stars & 3324 \\
\verb|D_logg| &  $\log(\textrm{cms}^{-2})$ & Dispersion in $\log g$ for neighborhood stars & 0.27 \\
\verb|M_logg| &  $\log(\textrm{cms}^{-2})$ & Median of $\log g$ for neighborhood stars & 4.81 \\
\verb|D_Ab_Mag| & G-mag & Dispersion in absolute mag of neighborhood stars & 1.05 \\
\verb|M_Ab_Mag| & G-mag & Median of absolute mag of neighborhood stars & 15.49 \\
\verb|N.S.I| & - & Neighborhood similarity index & 0.527 \\
\verb|NSI_std| & - & Uncertainty in neighborhood similarity index & 0.001 \\
\hline
\end{tabular}
\caption{Description of the content of the table of 84 HZS (with HD 165155 as an example). `Neighborhood stars' refers to the stars present within the 10~pc environment of the respective HZS. The entire table is available in machine-readable format.}
    \label{tab:1}
\end{table*}

The discovery of numerous extrasolar planets has revealed a diverse array of stellar and planetary characteristics, making systematic comparisons crucial for evaluating habitability and assessing the potential for life beyond our solar system. For example, Sun-like stars are more likely to host stable habitable environments for their planetary systems due to the moderate nature of their stellar activity \citep{loeb2016relative,Lingam_2017,Lingam_2017b,haqq2018we}. Additionally, the overall stellar environment of the solar system (e.g., location in the galaxy, neighborhood stellar spectral types, density and dispersion velocity) appears conducive to the long-term maintenance of habitable conditions \citep{Torres_2013,spitoni2017galactic,spinelli2021best,Spinelli_2023} and may serve as a valuable reference for comparison \citep{reyle202110,Gaia_10pc_Update}. Furthermore, since the Sun has played a crucial role in the evolution and maintenance of life on Earth, assessing the similarity of other planet-hosting stars to the Sun is of paramount importance.

The similarity index is a numerical metric used to quantify the likeness or resemblance between objects or systems sharing specific properties. This concept has been previously applied to assess the resemblance of known extrasolar planets to Earth by comparing their mass, radius, and surface temperature \citep{cha2007comprehensive,Schulze_2011}.
In this study, we employ the concept of similarity index to assess the resemblance and dissimilarity between 84 habitable systems and their stellar environments. Specifically, we utilize two distinct similarity indices:
\begin{enumerate}
    \item \textbf{Solar Similarity Index (SSI):} This index allows for a comparison of the properties of our Solar System with those of corresponding HZS.
    \item \textbf{Neighborhood Similarity Index (NSI):} This index facilitates a comparison of the properties of stars in  the 10~pc volume around Solar System and the neighborhood stars of the HZS. 
\end{enumerate}

We define each similarity index as:
\begin{equation}\label{eq:8}
    SSI, NSI = 1 - \sqrt{\frac{1}{n}\sum_{i=1}^{n} \left(\frac{P_i - P_{i \odot}}{P_i + P_{i \odot}}\right) ^2}
\end{equation}
where $P_i$ refers to the value of the $i$-th stellar parameter chosen for comparison  while $P_{i\odot}$ is the corresponding value for the Sun or the solar system and $n$ denotes the number of parameters considered. 

To calculate the SSI using Equation~\eqref{eq:8} we selected four parameters, namely, stellar multiplicity, effective temperature, surface gravity and absolute magnitude. Likewise, to calculate the NSI for each HZS, we chose eight parameters for the ensemble of stars within the 10~pc region. These parameters include the dispersion velocity of stars ($\sigma_v$), the median and standard deviation of temperature ($\langle \textrm{T}_{\textrm{eff}}\rangle$ and $\sigma_{T_{\text{eff}}}$), the median and standard deviation of absolute magnitude ($\langle \textrm{mag} \rangle$ and $\sigma_{mag}$), the median and standard deviation of $\log g$ ($\langle\log g\rangle$ and $\sigma_{\log g}$), and finally the neighborhood star count.

The SSI was calculated directly from Equation~\eqref{eq:8} for each HZ star, using stellar parameters taken from NEA. For the NSI of a stellar environment, we used the bootstrap method to generate 100,000 neighborhood configurations. We then derived the NSI value from the median of this sampling distribution and its associated uncertainties from the standard deviation (see Appendix~\ref{derive}).

The plot of NSI versus SSI in Figure~\ref{ssi-nsi} shows the comparison of different  HZS with solar system and stars in their respective 10~pc neighborhood. The HZS within the blue circle represent a similarity greater than 0.75 while those with more than 0.50 similarity fall within the orange circle. Overall, we note that NSI is $>0.75$ for most HZS implying a high degree of similarity between the stellar environment of the Sun and the HZ stars. Also, the HZS nearer to the solar neighborhood tend to have a higher NSI due to the presence of a similar population of stars. The differences however, tend to grow with an increase in distance as evident from the vertical color gradient seen in Figure~\ref{ssi-nsi}. On the other hand, SSI values have a larger spread primarily due to the different spectral types of stars in the HZ sample.

More specifically, in Figure~\ref{ssi-nsi} we note that HD~40307, a K2.5V dwarf star with T$\textrm{eff}\approx 5000$~K, located 13~pc from the Sun and hosting a 7.1~M$_\oplus$  habitable zone planet HD~40307g \citep{2009A&A...493..639M,2013A&A...549A..48T,10.1093/mnras/stu555}, has the highest overall similarity index among the 84~HZS. This system has an NSI of 0.94 and SSI of 0.92 and 5 planets discovered in its system. The NSI of Proxima Centauri is highest because its 10~pc volume significantly overlaps with the stars in the 10~pc neighborhood of the Sun. However, Proxima Centauri is a red-dwarf star (T$\textrm{eff}\approx 3000$~K) which is cooler and smaller than the Sun, and it is also part of a multiple star system. These differences contribute to its low SSI seen in Figure~\ref{ssi-nsi}. Conversely, HD~165155 is a G-type HZ star with a high SSI but low NSI. This is largely due to its densely populated surrounding, containing nearly 10,000 stars, in stark contrast to 315 stars in the 10~pc solar neighborhood. 

Exoplanet demographic studies have shown that the solar system is somewhat uncommon in terms of both its planetary properties such as mass, radius, orbital period, eccentricity and distribution in the system, as well as the stellar properties of the Sun \citep{Martin_2015, win15, zhu21}. Many studies have tried to find stars that closely resemble the Sun particularly in terms of fundamental stellar parameters, activity, rotation rate and elemental abundances \citep{solar_twins_2009,solar_twins_2015,solar_twins_2016}. Our current treatment of the SSI is somewhat simplistic. It is not intended to capture all the nuances and subtleties required to establish or refute the uniqueness of the Sun or the solar system among the known exoplanetary systems. Settling this question -- one way or another -- would require more data and improved characterization of host stars, which future ground-based and space missions will hopefully provide.
  
\begin{figure}[h]
\centering
\includegraphics[width=1.\columnwidth]{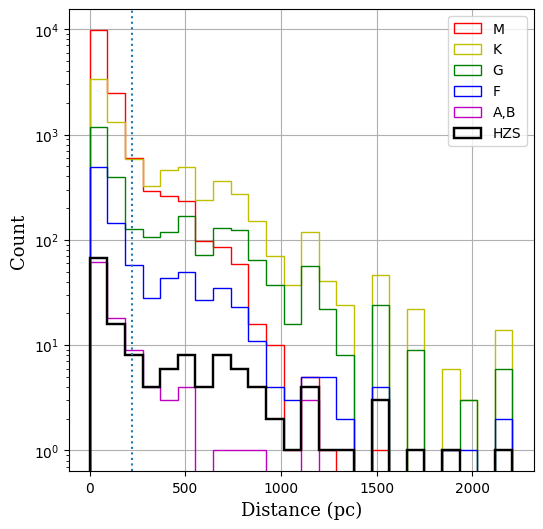}
\caption{Distribution of GAIA-detected neighborhood stars and HZS. The source count is a function of spectral type and distance from the Sun. Eighty four out of 146 HZS studied in this work lie within a distance of 220~pc (marked by a vertical dotted line) from the Sun. }
\label{dist_comp}
\end{figure}

\section{Completeness of GAIA data}\label{exhaust}

\begin{figure*}[!t]
    \centering
    \includegraphics[width=1.5\columnwidth]{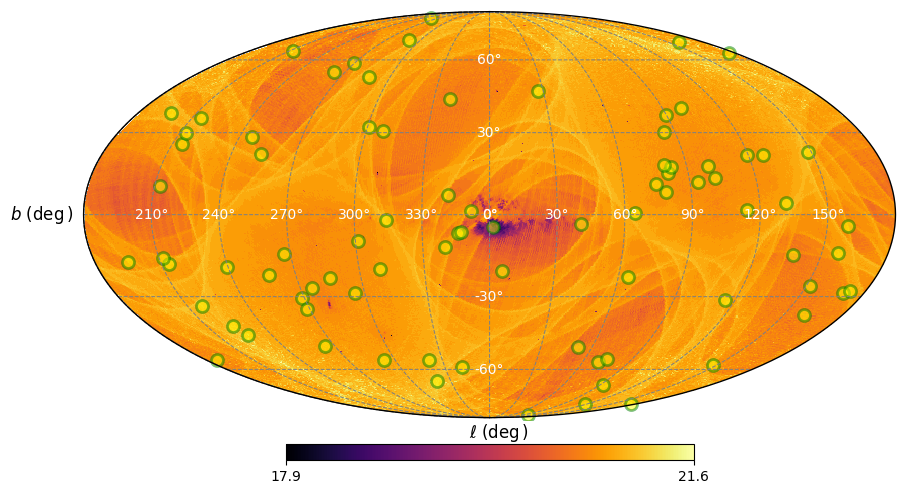}
    \caption{The 95\% completeness map of the Gaia-scanned region in Galactic coordinates ($l$ and $b$), generated using the  \texttt{gaiaunlimited} package \citep{complete}.  The color scale indicates the faintest G magnitude at which the 95\%  completeness threshold is achieved. Our sample of 84 HZS (green circles) has been overlaid on the map to visually depict the completeness of their respective neighborhoods.}
    \label{fig:comp_95}
\end{figure*}

Instrument sensitivities limit the detection of stars, as well as the measurement of stellar properties and the quantities derived from them \citep{Gaia_apsis_II}. Therefore, an instrument is capable of detecting and characterize only a certain fraction of the true number of objects (the ground-truth) -- this fraction is called completeness. The completeness of detections and the measurement/estimation of stellar astrophysical parameters (e.g., radial velocity, $T_{\textrm{eff}}$, etc) in \textit{Gaia DR3}, depends on various factors which include source crowding in the field of observation and scanning law (number of visits \textit{Gaia} made to that patch of the sky) \citep{complete}. This affects the accuracy of our results outlined in the previous section. In the following subsections, we discuss the completeness of our sample and its impact on threat assessment from stellar encounters and the interpretation of similarity indices.  

\subsection{Detection Completeness}

Stellar encounters heavily depend on the density of the stellar neighborhood; the number of stars is one of the parameters required to calculate the NSI. \textit{Gaia} is sensitive to detect objects with apparent magnitudes ranging from $G\approx3$ to $G\approx21$ \citep{eva2018}. Based on the simplified assumption that the solar neighborhood is a good representative of the demographics of stars within $\sim250$~pc, we can conclude that bright stars constitute a small fraction of the population whereas low-mass M-dwarfs and ultra-cool dwarfs constitute majority of the population \citep{gaia_10pc}. Figure \ref{dist_comp} shows the demographics of the {\it Gaia} detected 10~pc sample of neighborhood stars as a function of spectral types and distance from the Sun. At closer distances, the detection of most stars is nearly complete. The detection sensitivity begins to decline as the distance increases; however, this decline is faster for late spectral-type stars than for early spectral-type stars. Although T5 dwarfs have been detected in our sample up to 20~pc and the faintest source detected at 240 pc is an M8 dwarf, this doesn't guarantee a complete detection of similar sources. A single detection may be attributed to {\it Gaia}'s large number of visits to the corresponding patch of sky or to a relatively darker background \citep{complete}. In reality, there might be many more M-dwarfs and brown dwarfs in a given region than the few detections suggest.

The completeness of \textit{Gaia DR3} was derived by \citet{complete} by comparing detections of \textit{Gaia DR3} against that of DECaPS\footnote{Dark Energy Camera Plane Survey (DECaPS) was a ground-based deep survey of the southern galactic plane.} \citep{decaps}. \textit{Gaia} has a complex ``scanning law", and its performance varies significantly between sparse and densely populated regions of the sky. Therefore, the number of useful visits, and hence the total integration time of \textit{Gaia}, is not the same throughout the sky. These factors result in anisotropy of the median magnitude of stars detected by \textit{Gaia} in a particular patch of sky. \citet{complete} use the median magnitude of stars\footnote{Median magnitude of stars with useful visits (the \textit{Gaia} parameter, \texttt{astrometric\_matched\_transits}) $\gtrsim 10$, referred to in \citet{complete} as $M_{10}$. We refer to it as M here for simplicity} ($M$) to account for the anisotropic selection biases of \textit{Gaia} while empirically modeling its completeness. Figure \ref{fig:comp_95} shows the sky distribution of the faintest magnitude of stars in the Galactic reference system, to which \textit{Gaia} is at least 95\% complete. This map was generated using the \texttt{gaiaunlimited} package developed by \cite{complete}. The overlaid green points show the positions of 84 HZS that we analyze. This figure provides a qualitative description of the sky distribution of the apparent magnitude to which a particular HZS's neighborhood (fixed M) is 95\% complete.

DECaPS is a deep survey with a small footprint (~6.5\% sky coverage). Also, it is not complete to the faintest brown dwarfs and therefore, does not represent the ground truth. Although it can provide a good reference to estimate \textit{Gaia's} overall completeness, this method is not entirely reliable for computing the true counts for neighborhoods of individual HZS. We therefore, have used a selection function\footnote{Empirical model for \textit{Gaia} completeness: a sigmoid function with M and apparent magnitude as its parameters}, (see section 2.3 of \cite{complete}) to estimate  the completeness as a function of spectral type of stars in HZS neighborhoods. 

\begin{figure}[t!]
\centering
\includegraphics[width=1.\columnwidth]{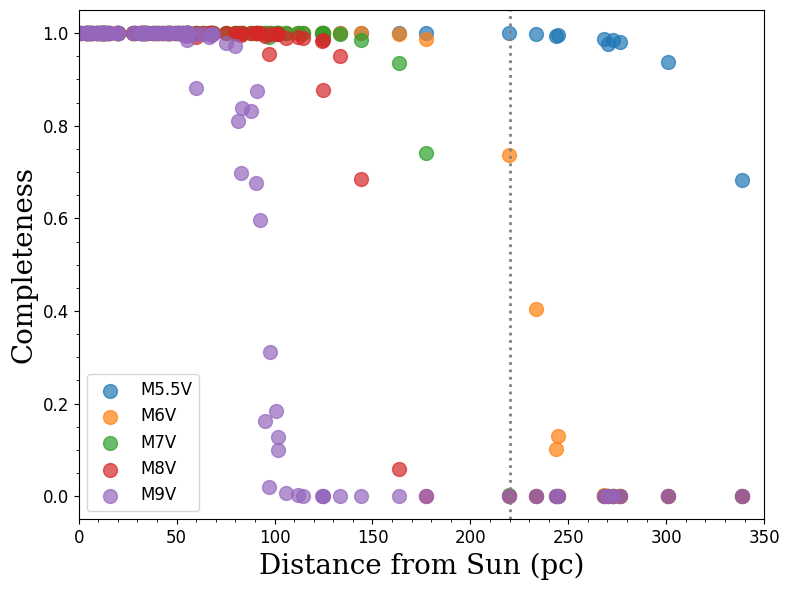}
\caption{{\it Gaia} completeness of neighborhood stars as a function of distance. Different colors represent different spectral types. Each point corresponds to a HZ system.}
\label{gaia_comp}
\end{figure}

For a particular spectral type, (i.e., certain absolute G magnitude) apparent magnitude increases with distance. For every HZ system, (different M: varying galactic co-ordinates) we compute the completeness using the selection function developed by \cite{complete}. Figure \ref{gaia_comp} shows how the completeness of neighborhood stars of different spectral types varies with distance. Notably, the detection of all stars up to the M-spectral type is nearly complete for HZS within $\sim$80~pc from the Sun. The location of the farthest exoplanetary system analyzed in our HZ sample, \textit{Kepler-296}, is marked by a vertical line drawn at 220~pc in Figure~\ref{gaia_comp}. We note that the stellar neighborhood of \textit{Kepler-296} is almost complete for M5.5V type stars, $\sim$~70\%  complete for M6V type, whereas ultra-cool dwarfs (later than M6V types) have a significantly lower likelihood of being detected.

\subsection{Completeness of Astrophysical Parameters}

The three main instruments onboard {\it Gaia} include an astrometric instrument for precise stellar position and parallax measurements, a broadband photometer for measuring stellar brightness, and a radial velocity spectrometer to determine the velocity of stars along the line of sight. The radial velocity spectra (RVS) is also used to estimate parameters like T$_\textrm{eff}$ and $\log g$. The photometric instrument provides the color information along with apparent magnitude, T$_\textrm{eff}$ and $\log g$. \cite{bla23, Gaia_apsis_I} and \cite{Gaia_apsis_II} provide a detailed discussion on {\it Gaia}'s estimates and accuracy of astrophysical parameters. Generally, the stellar parameters obtained using RVS are more accurate than those obtained using photometry. {\it Gaia} uses three methods to estimate T$_\textrm{eff}$ and $\log g$: Generalized Stellar Parametrizer (GSP), Extended Stellar Parametrizer (ESP), and Multiple Star Classifier (MSC). We obtained the stellar parameters of neighborhood stars from {\it Gaia} by following the priority order (from highest to lowest): 
GSP-Spec, GSP-Phot, ESP-HS, ESP-UCD, MSC-1, MSC-2\footnote{Spec: Spectroscopy, Phot: Photometry, HS: Hot Stars, UCD: Ultra-cool Dwarfs}.  GSP-Spec mainly operates on stars with $G\lesssim 15$, GSP-Phot on $G\lesssim 19$, MSC on $G\lesssim18.25$, and ESP-HS and ESP-UCD operate on  hot stars and ultra-cool dwarfs, respectively \citep{bla23, Gaia_apsis_I,Gaia_apsis_II}.

\begin{figure}[htpb]
\centering
\includegraphics[width=1.\columnwidth]{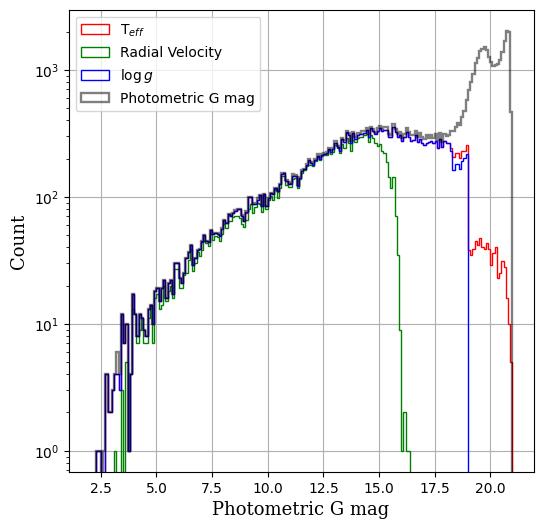}
\caption{Distribution of photometric mean G magnitude for objects with available $T_{\textrm{eff}}$, $\log g$ and radial velocity in \textit{Gaia}.}
\label{mag_comp}
\end{figure}

Figure \ref{mag_comp} illustrates the completeness  of various astrophysical parameters obtained from {\it Gaia} as a function of apparent $G$ magnitude of stars. Notably, the data for most parameters is complete up to $ G\approx 15$. Beyond that the completeness differs for each parameter. For example, the complete radial velocity data is only available for stars with $G \lesssim 15$, whereas  T$_\textrm{eff}$ and $\log g$ data is available for the majority of stars detected up to  $G \lesssim 17.5$. These magnitude limits are further described in \cite{Gaia_apsis_II}.

The foregoing discussions demonstrate that our 10~pc neighborhood dataset is incomplete both in terms of source detections and the availability of astrophysical parameters from {\it Gaia}. This incompleteness directly impacts the results presented in the section \ref{sec3}. For example, a fraction of low-mass faint stars at larger distances would remain undetected, leading to an underestimation of stellar density in the 10~pc region. Consequently, the estimated stellar encounter rates represent a lower limit of the actual values. Likewise, the completeness of astrophysical parameters is biased towards the brighter stars. This indicates that the median values of stellar parameters such as T$_\textrm{eff}$, $\log g$ derived from their respective bootstrap distributions are overestimated, while the neighborhood star count $n$, and absolute magnitude $M_G$ are underestimated. This introduces a completeness dependent bias in the estimation of NSI. Although this bias in NSI is minimized by restricting our analysis to the 84~HZS closest to the Sun, it is not entirely eliminated.

\section{Summary and Conclusion}\label{sec5}
The quest to find habitable planets is a key area of study in exoplanet research. More than 150 confirmed detections of planets in the habitable zones of various stars have been made. The origin and stability of habitable conditions depend on the star-planet system's location in the Milky Way and are influenced by factors such as high-energy radiation from supernovae, the presence of heavy elements, and the epoch of planet formation. High rates of catastrophic events in dense stellar regions can hinder the long-term evolution and survival of habitability.
For a planet to remain habitable, it must retain its atmosphere, be shielded from harmful radiation, and maintain a stable orbit within its habitable zone without being perturbed by other gravitational influences. The astrophysical impacts of stellar environment is a ``low-probability, high-consequence" scenario for the continuation of habitability of exoplanets. Even a single disruptive event of this kind, though less likely, could significantly impact the planet's habitability.

In this study, we focused on the 10~pc  neighborhood around known HZ systems to investigate the potential threats to their  habitability from nearby stellar encounters and supernovae. To accomplish these goals, we analyzed the astrometric, photometric, and spectroscopic data of these environments using {\it Gaia DR3}  and Hipparcos. We used a bootstrap approach to construct the 10~pc neighborhoods of HZ systems, discussed the influence of stellar environments on habitability and highlighted the limitations due to incomplete {\it Gaia} data. We also developed two metrics, the Solar Similarity Index and the Neighborhood Similarity Index, to compare the properties of the 10 pc environments of HZ systems with the 10~pc environment of our solar system. 

Out of the 84~HZS studied, 3 systems have a stellar density of $\geq 0.4 \textrm{pc}^{-3}$. Among these, only one system, HD~165155, has an encounter rate of $\geq 1$ in 5~Gyr period, increasing the likelihood of perturbation of planetary orbits during the star's main sequence evolution. We found a high-mass star ($>$8~M$_\odot$) within the 10~pc neighborhoods of each of the two HZS, namely TOI-1227 and HD~48265. These high-mass stars are potential progenitors for supernova explosions. Energy released from these stellar explosions can deposit a high fluence of harmful radiations on distant planets (or their exomoons), stripping off their ozone layer and rendering them uninhabitable. Upon comparing the 10~pc solar and stellar neighborhoods, we find that the stellar environments of the majority of HZS exhibit a high degree of similarity (NSI$>0.75$) to the solar neighborhood. Due to the diverse spectral types of HZ planet-hosting stars, when we compare them with the Sun, we get a wide range of SSI values. 

Finally, we discussed the possible limitations of this study due to the incompleteness of \textit{Gaia} data. We find that our sample of different HZ system's neighborhood stars is complete to early M-type stars. Incompleteness starts to plague our sample for ultra-cool dwarfs (later than M6V) at $\sim220~pc$. We also show that instruments onboard \textit{Gaia} are biased to brighter stars, \textit{Gaia DR3} completely catalogs all astrophysical parameters only till $G\lesssim15$. Therefore, by volume-limiting our sample to 84 HZS within 220~pc, we partially mitigated the uncertainties injected into our analysis due to \textit{Gaia}'s completeness bias. Because of \textit{Gaia} incompleteness, computed stellar encounter probabilities are lower limits and comparison between solar system and the exo-HZS is not accurate. However, our analysis provides a preliminary basis for characterizing the stellar environments of exoplanetary systems and warrants the need for more complete surveys. 

From the habitability perspective, investigating the local stellar environments of planet-hosting stars is an interesting and challenging problem that will benefit from more complete and accurate data. The forthcoming \textit{Gaia Data Release 4} (DR4) promises enhanced completeness and precision in estimating the astrophysical parameters of stars, which will improve our capability to fully characterize the 10~pc neighborhood of HZ stars. Additionally, future deep-sky surveys will further enhance our ability to investigate the stellar neighborhoods of planet-hosting stars at greater distances.

\section{Acknowledgement}
This work has made use of data from the European Space Agency (ESA) mission
{\it Gaia} (\url{https://www.cosmos.esa.int/gaia}), processed by the {\it Gaia}
Data Processing and Analysis Consortium (DPAC,
\url{https://www.cosmos.esa.int/web/gaia/dpac/consortium}). Funding for the DPAC
has been provided by national institutions, in particular the institutions
participating in the {\it Gaia} Multilateral Agreement.
This research has made use of data obtained through the High Energy Astrophysics Science Archive Research Center Online Service, provided by the NASA/Goddard Space Flight Center.
This work has also referred the NASA Exoplanet Archive and the Habitable Zone Catalog \citep{2012PASP..124..323K}. This research has made use of the Exoplanet Orbit Database
and the Exoplanet Data Explorer at exoplanets.org. Tisyagupta would also like to thank the Indian Institute of Astrophysics, for providing the necessary support during this work. Finally,  authors sincerely thank the anonymous reviewer for their valuable comments and suggestions, which have helped us improve the quality of this manuscript.

\textit{Software:} NumPy \citep{harris2020array}, PyAstronomy\footnote{\url{https://github.com/sczesla/PyAstronomy}} \citep{pya}, Astroquery \citep{2019AJ....157...98G}, Matplotlib \citep{Hunter:2007}.
\bibliographystyle{aasjournal}
\bibliography{biblio}

\begin{thebibliography}{}
\expandafter\ifx\csname natexlab\endcsname\relax\def\natexlab#1{#1}\fi
\providecommand{\url}[1]{\href{#1}{#1}}
\providecommand{\dodoi}[1]{doi:~\href{http://doi.org/#1}{\nolinkurl{#1}}}
\providecommand{\doeprint}[1]{\href{http://ascl.net/#1}{\nolinkurl{http://ascl.net/#1}}}
\providecommand{\doarXiv}[1]{\href{https://arxiv.org/abs/#1}{\nolinkurl{https://arxiv.org/abs/#1}}}

\bibitem[{{Airapetian} {et~al.}(2017){Airapetian}, {Glocer}, {Khazanov}, {Loyd}, {France}, {Sojka}, {Danchi}, \& {Liemohn}}]{atmosphere_loss}
{Airapetian}, V.~S., {Glocer}, A., {Khazanov}, G.~V., {et~al.} 2017, \apjl, 836, L3, \dodoi{10.3847/2041-8213/836/1/L3}

\bibitem[{Akeson {et~al.}(2013)Akeson, Chen, Ciardi, Crane, Good, Harbut, Jackson, Kane, Laity, Leifer, Lynn, McElroy, Papin, Plavchan, Ramírez, Rey, von Braun, Wittman, Abajian, Ali, Beichman, Beekley, Berriman, Berukoff, Bryden, Chan, Groom, Lau, Payne, Regelson, Saucedo, Schmitz, Stauffer, Wyatt, \& Zhang}]{Akeson_2013}
Akeson, R.~L., Chen, X., Ciardi, D., {et~al.} 2013, Publications of the Astronomical Society of the Pacific, 125, 989, \dodoi{10.1086/672273}

\bibitem[{{Banerjee} {et~al.}(2024){Banerjee}, {Narang}, {Manoj}, {Henning}, {Tyagi}, {Surya}, {Nayak}, \& {Tripathi}}]{ban24}
{Banerjee}, B., {Narang}, M., {Manoj}, P., {et~al.} 2024, \aj, 168, 7, \dodoi{10.3847/1538-3881/ad429f}

\bibitem[{{Baxter} {et~al.}(2018){Baxter}, {Blake}, \& {Jain}}]{Oort_Baxter}
{Baxter}, E.~J., {Blake}, C.~H., \& {Jain}, B. 2018, \aj, 156, 243, \dodoi{10.3847/1538-3881/aae64e}

\bibitem[{{Bojnordi Arbab} \& {Rahvar}(2021)}]{Arbab_2021}
{Bojnordi Arbab}, B., \& {Rahvar}, S. 2021, International Journal of Modern Physics D, 30, 2150063, \dodoi{10.1142/S0218271821500632}

\bibitem[{Brasser {et~al.}(2014)Brasser, Ida, \& Kokubo}]{10.1093/mnras/stu555}
Brasser, R., Ida, S., \& Kokubo, E. 2014, Monthly Notices of the Royal Astronomical Society, 440, 3685, \dodoi{10.1093/mnras/stu555}

\bibitem[{{Cai} {et~al.}(2017){Cai}, {Kouwenhoven}, {Portegies Zwart}, \& {Spurzem}}]{Cai_2017}
{Cai}, M.~X., {Kouwenhoven}, M.~B.~N., {Portegies Zwart}, S.~F., \& {Spurzem}, R. 2017, \mnras, 470, 4337, \dodoi{10.1093/mnras/stx1464}

\bibitem[{{Cantat-Gaudin, Tristan} {et~al.}(2023){Cantat-Gaudin, Tristan}, {Fouesneau, Morgan}, {Rix, Hans-Walter}, {Brown, Anthony G. A.}, {Castro-Ginard, Alfred}, {Kostrzewa-Rutkowska, Zuzanna}, {Drimmel, Ronald}, {Hogg, David W.}, {Casey, Andrew R.}, {Khanna, Shourya}, {Oh, Semyeong}, {Price-Whelan, Adrian M.}, {Belokurov, Vasily}, {Saydjari, Andrew K.}, \& {Green, G.}}]{complete}
{Cantat-Gaudin, Tristan}, {Fouesneau, Morgan}, {Rix, Hans-Walter}, {et~al.} 2023, A\&A, 669, A55, \dodoi{10.1051/0004-6361/202244784}

\bibitem[{Cha(2007)}]{cha2007comprehensive}
Cha, S.-H. 2007, City, 1, 1

\bibitem[{{Co{\c{s}}kuno{\v{g}}lu} {et~al.}(2011){Co{\c{s}}kuno{\v{g}}lu}, {Ak}, {Bilir}, {Karaali}, {Yaz}, {Gilmore}, {Seabroke}, {Bienaym{\'e}}, {Bland-Hawthorn}, {Campbell}, {Freeman}, {Gibson}, {Grebel}, {Munari}, {Navarro}, {Parker}, {Siebert}, {Siviero}, {Steinmetz}, {Watson}, {Wyse}, \& {Zwitter}}]{UVW_LSR}
{Co{\c{s}}kuno{\v{g}}lu}, B., {Ak}, S., {Bilir}, S., {et~al.} 2011, \mnras, 412, 1237, \dodoi{10.1111/j.1365-2966.2010.17983.x}

\bibitem[{{Creevey, O. L.} {et~al.}(2023){Creevey, O. L.}, {Sordo, R.}, {Pailler, F.}, {Fr{\'e}mat, Y.}, {Heiter, U.}, {Th{\'e}venin, F.}, {Andrae, R.}, {Fouesneau, M.}, {Lobel, A.}, {Bailer-Jones, C. A. L.}, {Garabato, D.}, {Bellas-Velidis, I.}, {Brugaletta, E.}, {Lorca, A.}, {Ordenovic, C.}, {Palicio, P. A.}, {Sarro, L. M.}, {Delchambre, L.}, {Drimmel, R.}, {Rybizki, J.}, {Torralba Elipe, G.}, {Korn, A. J.}, {Recio-Blanco, A.}, {Schultheis, M. S.}, {De Angeli, F.}, {Montegriffo, P.}, {Abreu Aramburu, A.}, {Accart, S.}, {{\'A}lvarez, M. A.}, {Bakker, J.}, {Brouillet, N.}, {Burlacu, A.}, {Carballo, R.}, {Casamiquela, L.}, {Chiavassa, A.}, {Contursi, G.}, {Cooper, W. J.}, {Dafonte, C.}, {Dapergolas, A.}, {de Laverny, P.}, {Dharmawardena, T. E.}, {Edvardsson, B.}, {Le Fustec, Y.}, {García-Lario, P.}, {García-Torres, M.}, {Gomez, A.}, {Gonz{\'a}lez-Santamaría, I.}, {Hatzidimitriou, D.}, {Jean-Antoine Piccolo, A.}, {Kontiza, M.}, {Kordopatis, G.}, {Lanzafame, A. C.}, {Lebreton, Y.}, {Licata, E. L.},
  {Lindstrøm, H. E. P.}, {Livanou, E.}, {Magdaleno Romeo, A.}, {Manteiga, M.}, {Marocco, F.}, {Marshall, D. J.}, {Mary, N.}, {Nicolas, C.}, {Pallas-Quintela, L.}, {Panem, C.}, {Pichon, B.}, {Poggio, E.}, {Riclet, F.}, {Robin, C.}, {Santoveña, R.}, {Silvelo, A.}, {Slezak, I.}, {Smart, R. L.}, {Soubiran, C.}, {Süveges, M.}, {Ulla, A.}, {Utrilla, E.}, {Vallenari, A.}, {Zhao, H.}, {Zorec, J.}, {Barrado, D.}, {Bijaoui, A.}, {Bouret, J.-C.}, {Blomme, R.}, {Brott, I.}, {Cassisi, S.}, {Kochukhov, O.}, {Martayan, C.}, {Shulyak, D.}, \& {Silvester, J.}}]{Gaia_apsis_I}
{Creevey, O. L.}, {Sordo, R.}, {Pailler, F.}, {et~al.} 2023, A\&A, 674, A26, \dodoi{10.1051/0004-6361/202243688}

\bibitem[{{Czesla} {et~al.}(2019){Czesla}, {Schr{\"o}ter}, {Schneider}, {Huber}, {Pfeifer}, {Andreasen}, \& {Zechmeister}}]{pya}
{Czesla}, S., {Schr{\"o}ter}, S., {Schneider}, C.~P., {et~al.} 2019, {PyA: Python astronomy-related packages}.
\newblock \doeprint{1906.010}

\bibitem[{{Datson} {et~al.}(2015){Datson}, {Flynn}, \& {Portinari}}]{solar_twins_2015}
{Datson}, J., {Flynn}, C., \& {Portinari}, L. 2015, \aap, 574, A124, \dodoi{10.1051/0004-6361/201425000}

\bibitem[{{Davari} {et~al.}(2022){Davari}, {Capuzzo-Dolcetta}, \& {Spurzem}}]{dav22}
{Davari}, N., {Capuzzo-Dolcetta}, R., \& {Spurzem}, R. 2022, \mnras, 513, 90, \dodoi{10.1093/mnras/stac462}

\bibitem[{Davison {et~al.}(2003)Davison, Hinkley, \& Young}]{Bootstrap_Recent_Devs}
Davison, A.~C., Hinkley, D.~V., \& Young, G.~A. 2003, Statistical Science, 18, 141 , \dodoi{10.1214/ss/1063994969}

\bibitem[{{de Juan Ovelar} {et~al.}(2012){de Juan Ovelar}, {Kruijssen}, {Bressert}, {Testi}, {Bastian}, \& {C{\'a}novas}}]{Ovelar_2012}
{de Juan Ovelar}, M., {Kruijssen}, J.~M.~D., {Bressert}, E., {et~al.} 2012, \aap, 546, L1, \dodoi{10.1051/0004-6361/201219627}

\bibitem[{{Evans} {et~al.}(2018){Evans}, {Riello}, {De Angeli}, {Carrasco}, {Montegriffo}, {Fabricius}, {Jordi}, {Palaversa}, {Diener}, {Busso}, {Cacciari}, {van Leeuwen}, {Burgess}, {Davidson}, {Harrison}, {Hodgkin}, {Pancino}, {Richards}, {Altavilla}, {Balaguer-N{\'u}{\~n}ez}, {Barstow}, {Bellazzini}, {Brown}, {Castellani}, {Cocozza}, {De Luise}, {Delgado}, {Ducourant}, {Galleti}, {Gilmore}, {Giuffrida}, {Holl}, {Kewley}, {Koposov}, {Marinoni}, {Marrese}, {Osborne}, {Piersimoni}, {Portell}, {Pulone}, {Ragaini}, {Sanna}, {Terrett}, {Walton}, {Wevers}, \& {Wyrzykowski}}]{eva2018}
{Evans}, D.~W., {Riello}, M., {De Angeli}, F., {et~al.} 2018, \aap, 616, A4, \dodoi{10.1051/0004-6361/201832756}

\bibitem[{{Fouesneau, M.} {et~al.}(2023){Fouesneau, M.}, {Fr{\'e}mat, Y.}, {Andrae, R.}, {Korn, A. J.}, {Soubiran, C.}, {Kordopatis, G.}, {Vallenari, A.}, {Heiter, U.}, {Creevey, O. L.}, {Sarro, L. M.}, {de Laverny, P.}, {Lanzafame, A. C.}, {Lobel, A.}, {Sordo, R.}, {Rybizki, J.}, {Slezak, I.}, {{\'A}lvarez, M. A.}, {Drimmel, R.}, {Garabato, D.}, {Delchambre, L.}, {Bailer-Jones, C. A. L.}, {Hatzidimitriou, D.}, {Lorca, A.}, {Le Fustec, Y.}, {Pailler, F.}, {Mary, N.}, {Robin, C.}, {Utrilla, E.}, {Abreu Aramburu, A.}, {Bakker, J.}, {Bellas-Velidis, I.}, {Bijaoui, A.}, {Blomme, R.}, {Bouret, J.-C.}, {Brouillet, N.}, {Brugaletta, E.}, {Burlacu, A.}, {Carballo, R.}, {Casamiquela, L.}, {Chaoul, L.}, {Chiavassa, A.}, {Contursi, G.}, {Cooper, W. J.}, {Dafonte, C.}, {Demouchy, C.}, {Dharmawardena, T. E.}, {García-Lario, P.}, {García-Torres, M.}, {Gomez, A.}, {Gonz{\'a}lez-Santamaría, I.}, {Jean-Antoine Piccolo, A.}, {Kontizas, M.}, {Lebreton, Y.}, {Licata, E. L.}, {Lindstrøm, H. E. P.}, {Livanou, E.}, {Magdaleno
  Romeo, A.}, {Manteiga, M.}, {Marocco, F.}, {Martayan, C.}, {Marshall, D. J.}, {Nicolas, C.}, {Ordenovic, C.}, {Palicio, P. A.}, {Pallas-Quintela, L.}, {Pichon, B.}, {Poggio, E.}, {Recio-Blanco, A.}, {Riclet, F.}, {Santoveña, R.}, {Schultheis, M. S.}, {Segol, M.}, {Silvelo, A.}, {Smart, R. L.}, {Süveges, M.}, {Th{\'e}venin, F.}, {Torralba Elipe, G.}, {Ulla, A.}, {van Dillen, E.}, {Zhao, H.}, \& {Zorec, J.}}]{Gaia_apsis_II}
{Fouesneau, M.}, {Fr{\'e}mat, Y.}, {Andrae, R.}, {et~al.} 2023, A\&A, 674, A28, \dodoi{10.1051/0004-6361/202243919}

\bibitem[{Fujii {et~al.}(2018)Fujii, Angerhausen, Deitrick, Domagal-Goldman, Grenfell, Hori, Kane, Palle, Rauer, Siegler, Stapelfeldt, \& Stevenson}]{Fujii_2018}
Fujii, Y., Angerhausen, D., Deitrick, R., {et~al.} 2018, Astrobiology, 18, 739, \dodoi{10.1089/ast.2017.1733}

\bibitem[{{Gaia Collaboration} {et~al.}(2023{\natexlab{a}}){Gaia Collaboration}, {Vallenari}, {Brown}, {Prusti}, {de Bruijne}, {Arenou}, {Babusiaux}, {Biermann}, {Creevey}, {Ducourant}, {Evans}, {Eyer}, {Guerra}, {Hutton}, {Jordi}, {Klioner}, {Lammers}, {Lindegren}, {Luri}, {Mignard}, {Panem}, {Pourbaix}, {Randich}, {Sartoretti}, {Soubiran}, {Tanga}, {Walton}, {Bailer-Jones}, {Bastian}, {Drimmel}, {Jansen}, {Katz}, {Lattanzi}, {van Leeuwen}, {Bakker}, {Cacciari}, {Casta{\~n}eda}, {De Angeli}, {Fabricius}, {Fouesneau}, {Fr{\'e}mat}, {Galluccio}, {Guerrier}, {Heiter}, {Masana}, {Messineo}, {Mowlavi}, {Nicolas}, {Nienartowicz}, {Pailler}, {Panuzzo}, {Riclet}, {Roux}, {Seabroke}, {Sordo}, {Th{\'e}venin}, {Gracia-Abril}, {Portell}, {Teyssier}, {Altmann}, {Andrae}, {Audard}, {Bellas-Velidis}, {Benson}, {Berthier}, {Blomme}, {Burgess}, {Busonero}, {Busso}, {C{\'a}novas}, {Carry}, {Cellino}, {Cheek}, {Clementini}, {Damerdji}, {Davidson}, {de Teodoro}, {Nu{\~n}ez Campos}, {Delchambre}, {Dell'Oro}, {Esquej},
  {Fern{\'a}ndez-Hern{\'a}ndez}, {Fraile}, {Garabato}, {Garc{\'\i}a-Lario}, {Gosset}, {Haigron}, {Halbwachs}, {Hambly}, {Harrison}, {Hern{\'a}ndez}, {Hestroffer}, {Hodgkin}, {Holl}, {Jan{\ss}en}, {Jevardat de Fombelle}, {Jordan}, {Krone-Martins}, {Lanzafame}, {L{\"o}ffler}, {Marchal}, {Marrese}, {Moitinho}, {Muinonen}, {Osborne}, {Pancino}, {Pauwels}, {Recio-Blanco}, {Reyl{\'e}}, {Riello}, {Rimoldini}, {Roegiers}, {Rybizki}, {Sarro}, {Siopis}, {Smith}, {Sozzetti}, {Utrilla}, {van Leeuwen}, {Abbas}, {{\'A}brah{\'a}m}, {Abreu Aramburu}, {Aerts}, {Aguado}, {Ajaj}, {Aldea-Montero}, {Altavilla}, {{\'A}lvarez}, {Alves}, {Anders}, {Anderson}, {Anglada Varela}, {Antoja}, {Baines}, {Baker}, {Balaguer-N{\'u}{\~n}ez}, {Balbinot}, {Balog}, {Barache}, {Barbato}, {Barros}, {Barstow}, {Bartolom{\'e}}, {Bassilana}, {Bauchet}, {Becciani}, {Bellazzini}, {Berihuete}, {Bernet}, {Bertone}, {Bianchi}, {Binnenfeld}, {Blanco-Cuaresma}, {Blazere}, {Boch}, {Bombrun}, {Bossini}, {Bouquillon}, {Bragaglia}, {Bramante}, {Breedt},
  {Bressan}, {Brouillet}, {Brugaletta}, {Bucciarelli}, {Burlacu}, {Butkevich}, {Buzzi}, {Caffau}, {Cancelliere}, {Cantat-Gaudin}, {Carballo}, {Carlucci}, {Carnerero}, {Carrasco}, {Casamiquela}, {Castellani}, {Castro-Ginard}, {Chaoul}, {Charlot}, {Chemin}, {Chiaramida}, {Chiavassa}, {Chornay}, {Comoretto}, {Contursi}, {Cooper}, {Cornez}, {Cowell}, {Crifo}, {Cropper}, {Crosta}, {Crowley}, {Dafonte}, {Dapergolas}, {David}, {David}, {de Laverny}, {De Luise}, {De March}, {De Ridder}, {de Souza}, {de Torres}, {del Peloso}, {del Pozo}, {Delbo}, {Delgado}, {Delisle}, {Demouchy}, {Dharmawardena}, {Di Matteo}, {Diakite}, {Diener}, {Distefano}, {Dolding}, {Edvardsson}, {Enke}, {Fabre}, {Fabrizio}, {Faigler}, {Fedorets}, {Fernique}, {Fienga}, {Figueras}, {Fournier}, {Fouron}, {Fragkoudi}, {Gai}, {Garcia-Gutierrez}, {Garcia-Reinaldos}, {Garc{\'\i}a-Torres}, {Garofalo}, {Gavel}, {Gavras}, {Gerlach}, {Geyer}, {Giacobbe}, {Gilmore}, {Girona}, {Giuffrida}, {Gomel}, {Gomez}, {Gonz{\'a}lez-N{\'u}{\~n}ez},
  {Gonz{\'a}lez-Santamar{\'\i}a}, {Gonz{\'a}lez-Vidal}, {Granvik}, {Guillout}, {Guiraud}, {Guti{\'e}rrez-S{\'a}nchez}, {Guy}, {Hatzidimitriou}, {Hauser}, {Haywood}, {Helmer}, {Helmi}, {Sarmiento}, {Hidalgo}, {Hilger}, {H{\l}adczuk}, {Hobbs}, {Holland}, {Huckle}, {Jardine}, {Jasniewicz}, {Jean-Antoine Piccolo}, {Jim{\'e}nez-Arranz}, {Jorissen}, {Juaristi Campillo}, {Julbe}, {Karbevska}, {Kervella}, {Khanna}, {Kontizas}, {Kordopatis}, {Korn}, {K{\'o}sp{\'a}l}, {Kostrzewa-Rutkowska}, {Kruszy{\'n}ska}, {Kun}, {Laizeau}, {Lambert}, {Lanza}, {Lasne}, {Le Campion}, {Lebreton}, {Lebzelter}, {Leccia}, {Leclerc}, {Lecoeur-Taibi}, {Liao}, {Licata}, {Lindstr{\o}m}, {Lister}, {Livanou}, {Lobel}, {Lorca}, {Loup}, {Madrero Pardo}, {Magdaleno Romeo}, {Managau}, {Mann}, {Manteiga}, {Marchant}, {Marconi}, {Marcos}, {Marcos Santos}, {Mar{\'\i}n Pina}, {Marinoni}, {Marocco}, {Marshall}, {Martin Polo}, {Mart{\'\i}n-Fleitas}, {Marton}, {Mary}, {Masip}, {Massari}, {Mastrobuono-Battisti}, {Mazeh}, {McMillan}, {Messina}, {Michalik},
  {Millar}, {Mints}, {Molina}, {Molinaro}, {Moln{\'a}r}, {Monari}, {Mongui{\'o}}, {Montegriffo}, {Montero}, {Mor}, {Mora}, {Morbidelli}, {Morel}, {Morris}, {Muraveva}, {Murphy}, {Musella}, {Nagy}, {Noval}, {Oca{\~n}a}, {Ogden}, {Ordenovic}, {Osinde}, {Pagani}, {Pagano}, {Palaversa}, {Palicio}, {Pallas-Quintela}, {Panahi}, {Payne-Wardenaar}, {Pe{\~n}alosa Esteller}, {Penttil{\"a}}, {Pichon}, {Piersimoni}, {Pineau}, {Plachy}, {Plum}, {Poggio}, {Pr{\v{s}}a}, {Pulone}, {Racero}, {Ragaini}, {Rainer}, {Raiteri}, {Rambaux}, {Ramos}, {Ramos-Lerate}, {Re Fiorentin}, {Regibo}, {Richards}, {Rios Diaz}, {Ripepi}, {Riva}, {Rix}, {Rixon}, {Robichon}, {Robin}, {Robin}, {Roelens}, {Rogues}, {Rohrbasser}, {Romero-G{\'o}mez}, {Rowell}, {Royer}, {Ruz Mieres}, {Rybicki}, {Sadowski}, {S{\'a}ez N{\'u}{\~n}ez}, {Sagrist{\`a} Sell{\'e}s}, {Sahlmann}, {Salguero}, {Samaras}, {Sanchez Gimenez}, {Sanna}, {Santove{\~n}a}, {Sarasso}, {Schultheis}, {Sciacca}, {Segol}, {Segovia}, {S{\'e}gransan}, {Semeux}, {Shahaf}, {Siddiqui}, {Siebert},
  {Siltala}, {Silvelo}, {Slezak}, {Slezak}, {Smart}, {Snaith}, {Solano}, {Solitro}, {Souami}, {Souchay}, {Spagna}, {Spina}, {Spoto}, {Steele}, {Steidelm{\"u}ller}, {Stephenson}, {S{\"u}veges}, {Surdej}, {Szabados}, {Szegedi-Elek}, {Taris}, {Taylor}, {Teixeira}, {Tolomei}, {Tonello}, {Torra}, {Torra}, {Torralba Elipe}, {Trabucchi}, {Tsounis}, {Turon}, {Ulla}, {Unger}, {Vaillant}, {van Dillen}, {van Reeven}, {Vanel}, {Vecchiato}, {Viala}, {Vicente}, {Voutsinas}, {Weiler}, {Wevers}, {Wyrzykowski}, {Yoldas}, {Yvard}, {Zhao}, {Zorec}, {Zucker}, \& {Zwitter}}]{gai2023}
{Gaia Collaboration}, {Vallenari}, A., {Brown}, A.~G.~A., {et~al.} 2023{\natexlab{a}}, \aap, 674, A1, \dodoi{10.1051/0004-6361/202243940}

\bibitem[{{Gaia Collaboration} {et~al.}(2023{\natexlab{b}}){Gaia Collaboration}, {Creevey, O. L.}, {Sarro, L. M.}, {Lobel, A.}, {Pancino, E.}, {Andrae, R.}, {Smart, R. L.}, {Clementini, G.}, {Heiter, U.}, {Korn, A. J.}, {Fouesneau, M.}, {Fr{\'e}mat, Y.}, {De Angeli, F.}, {Vallenari, A.}, {Harrison, D. L.}, {Th{\'e}venin, F.}, {Reyl{\'e}, C.}, {Sordo, R.}, {Garofalo, A.}, {Brown, A. G. A.}, {Eyer, L.}, {Prusti, T.}, {de Bruijne, J. H. J.}, {Arenou, F.}, {Babusiaux, C.}, {Biermann, M.}, {Ducourant, C.}, {Evans, D. W.}, {Guerra, R.}, {Hutton, A.}, {Jordi, C.}, {Klioner, S. A.}, {Lammers, U. L.}, {Lindegren, L.}, {Luri, X.}, {Mignard, F.}, {Panem, C.}, {Pourbaix, D.}, {Randich, S.}, {Sartoretti, P.}, {Soubiran, C.}, {Tanga, P.}, {Walton, N. A.}, {Bailer-Jones, C. A. L.}, {Bastian, U.}, {Drimmel, R.}, {Jansen, F.}, {Katz, D.}, {Lattanzi, M. G.}, {van Leeuwen, F.}, {Bakker, J.}, {Cacciari, C.}, {Castañeda, J.}, {Fabricius, C.}, {Galluccio, L.}, {Guerrier, A.}, {Masana, E.}, {Messineo, R.}, {Mowlavi, N.}, {Nicolas,
  C.}, {Nienartowicz, K.}, {Pailler, F.}, {Panuzzo, P.}, {Riclet, F.}, {Roux, W.}, {Seabroke, G. M.}, {Gracia-Abril, G.}, {Portell, J.}, {Teyssier, D.}, {Altmann, M.}, {Audard, M.}, {Bellas-Velidis, I.}, {Benson, K.}, {Berthier, J.}, {Blomme, R.}, {Burgess, P. W.}, {Busonero, D.}, {Busso, G.}, {C{\'a}novas, H.}, {Carry, B.}, {Cellino, A.}, {Cheek, N.}, {Damerdji, Y.}, {Davidson, M.}, {de Teodoro, P.}, {Nuñez Campos, M.}, {Delchambre, L.}, {Dell’Oro, A.}, {Esquej, P.}, {Fern{\'a}ndez-Hern{\'a}ndez, J.}, {Fraile, E.}, {Garabato, D.}, {García-Lario, P.}, {Gosset, E.}, {Haigron, R.}, {Halbwachs, J.-L.}, {Hambly, N. C.}, {Hern{\'a}ndez, J.}, {Hestroffer, D.}, {Hodgkin, S. T.}, {Holl, B.}, {Janßen, K.}, {Jevardat de Fombelle, G.}, {Jordan, S.}, {Krone-Martins, A.}, {Lanzafame, A. C.}, {Löffler, W.}, {Marchal, O.}, {Marrese, P. M.}, {Moitinho, A.}, {Muinonen, K.}, {Osborne, P.}, {Pauwels, T.}, {Recio-Blanco, A.}, {Riello, M.}, {Rimoldini, L.}, {Roegiers, T.}, {Rybizki, J.}, {Siopis, C.}, {Smith, M.},
  {Sozzetti, A.}, {Utrilla, E.}, {van Leeuwen, M.}, {Abbas, U.}, {{\'A}brah{\'a}m, P.}, {Abreu Aramburu, A.}, {Aerts, C.}, {Aguado, J. J.}, {Ajaj, M.}, {Aldea-Montero, F.}, {Altavilla, G.}, {{\'A}lvarez, M. A.}, {Alves, J.}, {Anders, F.}, {Anderson, R. I.}, {Anglada Varela, E.}, {Antoja, T.}, {Baines, D.}, {Baker, S. G.}, {Balaguer-Núñez, L.}, {Balbinot, E.}, {Balog, Z.}, {Barache, C.}, {Barbato, D.}, {Barros, M.}, {Barstow, M. A.}, {Bartolom{\'e}, S.}, {Bassilana, J.-L.}, {Bauchet, N.}, {Becciani, U.}, {Bellazzini, M.}, {Berihuete, A.}, {Bernet, M.}, {Bertone, S.}, {Bianchi, L.}, {Binnenfeld, A.}, {Blanco-Cuaresma, S.}, {Boch, T.}, {Bombrun, A.}, {Bossini, D.}, {Bouquillon, S.}, {Bragaglia, A.}, {Bramante, L.}, {Breedt, E.}, {Bressan, A.}, {Brouillet, N.}, {Brugaletta, E.}, {Bucciarelli, B.}, {Burlacu, A.}, {Butkevich, A. G.}, {Buzzi, R.}, {Caffau, E.}, {Cancelliere, R.}, {Cantat-Gaudin, T.}, {Carballo, R.}, {Carlucci, T.}, {Carnerero, M. I.}, {Carrasco, J. M.}, {Casamiquela, L.}, {Castellani, M.},
  {Castro-Ginard, A.}, {Chaoul, L.}, {Charlot, P.}, {Chemin, L.}, {Chiaramida, V.}, {Chiavassa, A.}, {Chornay, N.}, {Comoretto, G.}, {Contursi, G.}, {Cooper, W. J.}, {Cornez, T.}, {Cowell, S.}, {Crifo, F.}, {Cropper, M.}, {Crosta, M.}, {Crowley, C.}, {Dafonte, C.}, {Dapergolas, A.}, {David, P.}, {de Laverny, P.}, {De Luise, F.}, {De March, R.}, {De Ridder, J.}, {de Souza, R.}, {de Torres, A.}, {del Peloso, E. F.}, {del Pozo, E.}, {Delbo, M.}, {Delgado, A.}, {Delisle, J.-B.}, {Demouchy, C.}, {Dharmawardena, T. E.}, {Di Matteo, P.}, {Diakite, S.}, {Diener, C.}, {Distefano, E.}, {Dolding, C.}, {Enke, H.}, {Fabre, C.}, {Fabrizio, M.}, {Faigler, S.}, {Fedorets, G.}, {Fernique, P.}, {Figueras, F.}, {Fournier, Y.}, {Fouron, C.}, {Fragkoudi, F.}, {Gai, M.}, {Garcia-Gutierrez, A.}, {Garcia-Reinaldos, M.}, {García-Torres, M.}, {Gavel, A.}, {Gavras, P.}, {Gerlach, E.}, {Geyer, R.}, {Giacobbe, P.}, {Gilmore, G.}, {Girona, S.}, {Giuffrida, G.}, {Gomel, R.}, {Gomez, A.}, {Gonz{\'a}lez-Núñez, J.},
  {Gonz{\'a}lez-Santamaría, I.}, {Gonz{\'a}lez-Vidal, J. J.}, {Granvik, M.}, {Guillout, P.}, {Guiraud, J.}, {Guti{\'e}rrez-S{\'a}nchez, R.}, {Guy, L. P.}, {Hatzidimitriou, D.}, {Hauser, M.}, {Haywood, M.}, {Helmer, A.}, {Helmi, A.}, {Hilger, T.}, {Sarmiento, M. H.}, {Hidalgo, S. L.}, {Hładczuk, N.}, {Hobbs, D.}, {Holland, G.}, {Huckle, H. E.}, {Jardine, K.}, {Jasniewicz, G.}, {Jean-Antoine Piccolo, A.}, {Jim{\'e}nez-Arranz, Ó.}, {Juaristi Campillo, J.}, {Julbe, F.}, {Karbevska, L.}, {Kervella, P.}, {Khanna, S.}, {Kordopatis, G.}, {Kósp{\'a}l, {\'A}}, {Kostrzewa-Rutkowska, Z.}, {Kruszyńska, K.}, {Kun, M.}, {Laizeau, P.}, {Lambert, S.}, {Lanza, A. F.}, {Lasne, Y.}, {Le Campion, J.-F.}, {Lebreton, Y.}, {Lebzelter, T.}, {Leccia, S.}, {Leclerc, N.}, {Lecoeur-Taibi, I.}, {Liao, S.}, {Licata, E. L.}, {Lindstrøm, H. E. P.}, {Lister, T. A.}, {Livanou, E.}, {Lorca, A.}, {Loup, C.}, {Madrero Pardo, P.}, {Magdaleno Romeo, A.}, {Managau, S.}, {Mann, R. G.}, {Manteiga, M.}, {Marchant, J. M.}, {Marconi, M.}, {Marcos,
  J.}, {Marcos Santos, M. M. S.}, {Marín Pina, D.}, {Marinoni, S.}, {Marocco, F.}, {Marshall, D. J.}, {Martin Polo, L.}, {Martín-Fleitas, J. M.}, {Marton, G.}, {Mary, N.}, {Masip, A.}, {Massari, D.}, {Mastrobuono-Battisti, A.}, {Mazeh, T.}, {McMillan, P. J.}, {Messina, S.}, {Michalik, D.}, {Millar, N. R.}, {Mints, A.}, {Molina, D.}, {Molinaro, R.}, {Moln{\'a}r, L.}, {Monari, G.}, {Monguió, M.}, {Montegriffo, P.}, {Montero, A.}, {Mor, R.}, {Mora, A.}, {Morbidelli, R.}, {Morel, T.}, {Morris, D.}, {Muraveva, T.}, {Murphy, C. P.}, {Musella, I.}, {Nagy, Z.}, {Noval, L.}, {Ocaña, F.}, {Ogden, A.}, {Ordenovic, C.}, {Osinde, J. O.}, {Pagani, C.}, {Pagano, I.}, {Palaversa, L.}, {Palicio, P. A.}, {Pallas-Quintela, L.}, {Panahi, A.}, {Payne-Wardenaar, S.}, {Peñalosa Esteller, X.}, {Penttilä, A.}, {Pichon, B.}, {Piersimoni, A. M.}, {Pineau, F.-X.}, {Plachy, E.}, {Plum, G.}, {Poggio, E.}, {Prša, A.}, {Pulone, L.}, {Racero, E.}, {Ragaini, S.}, {Rainer, M.}, {Raiteri, C. M.}, {Ramos, P.}, {Ramos-Lerate, M.}, {Re
  Fiorentin, P.}, {Regibo, S.}, {Richards, P. J.}, {Rios Diaz, C.}, {Ripepi, V.}, {Riva, A.}, {Rix, H.-W.}, {Rixon, G.}, {Robichon, N.}, {Robin, A. C.}, {Robin, C.}, {Roelens, M.}, {Rogues, H. R. O.}, {Rohrbasser, L.}, {Romero-Gómez, M.}, {Rowell, N.}, {Royer, F.}, {Ruz Mieres, D.}, {Rybicki, K. A.}, {Sadowski, G.}, {S{\'a}ez Núñez, A.}, {Sagristà Sell{\'e}s, A.}, {Sahlmann, J.}, {Salguero, E.}, {Samaras, N.}, {Sanchez Gimenez, V.}, {Sanna, N.}, {Santoveña, R.}, {Sarasso, M.}, {Schultheis, M.}, {Sciacca, E.}, {Segol, M.}, {Segovia, J. C.}, {S{\'e}gransan, D.}, {Semeux, D.}, {Shahaf, S.}, {Siddiqui, H. I.}, {Siebert, A.}, {Siltala, L.}, {Silvelo, A.}, {Slezak, E.}, {Slezak, I.}, {Snaith, O. N.}, {Solano, E.}, {Solitro, F.}, {Souami, D.}, {Souchay, J.}, {Spagna, A.}, {Spina, L.}, {Spoto, F.}, {Steele, I. A.}, {Steidelmüller, H.}, {Stephenson, C. A.}, {Süveges, M.}, {Surdej, J.}, {Szabados, L.}, {Szegedi-Elek, E.}, {Taris, F.}, {Taylor, M. B.}, {Teixeira, R.}, {Tolomei, L.}, {Tonello, N.}, {Torra, F.},
  {Torra, J.}, {Torralba Elipe, G.}, {Trabucchi, M.}, {Tsounis, A. T.}, {Turon, C.}, {Ulla, A.}, {Unger, N.}, {Vaillant, M. V.}, {van Dillen, E.}, {van Reeven, W.}, {Vanel, O.}, {Vecchiato, A.}, {Viala, Y.}, {Vicente, D.}, {Voutsinas, S.}, {Weiler, M.}, {Wevers, T.}, {Wyrzykowski, Ł.}, {Yoldas, A.}, {Yvard, P.}, {Zhao, H.}, {Zorec, J.}, {Zucker, S.}, \& {Zwitter, T.}}]{Gaia_apsis_III}
{Gaia Collaboration}, {Creevey, O. L.}, {Sarro, L. M.}, {et~al.} 2023{\natexlab{b}}, A\&A, 674, A39, \dodoi{10.1051/0004-6361/202243800}

\bibitem[{{Garraffo} {et~al.}(2017){Garraffo}, {Drake}, {Cohen}, {Alvarado-G{\'o}mez}, \& {Moschou}}]{magnet_plasma}
{Garraffo}, C., {Drake}, J.~J., {Cohen}, O., {Alvarado-G{\'o}mez}, J.~D., \& {Moschou}, S.~P. 2017, \apjl, 843, L33, \dodoi{10.3847/2041-8213/aa79ed}

\bibitem[{{Ginsburg} {et~al.}(2019){Ginsburg}, {Sip{\H{o}}cz}, {Brasseur}, {Cowperthwaite}, {Craig}, {Deil}, {Guillochon}, {Guzman}, {Liedtke}, {Lian Lim}, {Lockhart}, {Mommert}, {Morris}, {Norman}, {Parikh}, {Persson}, {Robitaille}, {Segovia}, {Singer}, {Tollerud}, {de Val-Borro}, {Valtchanov}, {Woillez}, {Astroquery Collaboration}, \& {a subset of astropy Collaboration}}]{2019AJ....157...98G}
{Ginsburg}, A., {Sip{\H{o}}cz}, B.~M., {Brasseur}, C.~E., {et~al.} 2019, \aj, 157, 98, \dodoi{10.3847/1538-3881/aafc33}

\bibitem[{Glaser {et~al.}(2020)Glaser, Hartnett, Desch, Unterborn, Anbar, Buessecker, Fisher, Glaser, Kane, Lisse, Millsaps, Neuer, O’Rourke, Santos, Walker, \& Zolotov}]{Glaser_2020}
Glaser, D.~M., Hartnett, H.~E., Desch, S.~J., {et~al.} 2020, The Astrophysical Journal, 893, 163, \dodoi{10.3847/1538-4357/ab822d}

\bibitem[{{Gonzalez} {et~al.}(2001){Gonzalez}, {Brownlee}, \& {Ward}}]{gon2001}
{Gonzalez}, G., {Brownlee}, D., \& {Ward}, P. 2001, \icarus, 152, 185, \dodoi{10.1006/icar.2001.6617}

\bibitem[{{Han} {et~al.}(2014){Han}, {Wang}, {Wright}, {Feng}, {Zhao}, {Fakhouri}, {Brown}, \& {Hancock}}]{Han_2014}
{Han}, E., {Wang}, S.~X., {Wright}, J.~T., {et~al.} 2014, \pasp, 126, 827, \dodoi{10.1086/678447}

\bibitem[{Haqq-Misra {et~al.}(2018)Haqq-Misra, Kopparapu, \& Wolf}]{haqq2018we}
Haqq-Misra, J., Kopparapu, R.~K., \& Wolf, E.~T. 2018, International Journal of Astrobiology, 17, 77

\bibitem[{Harris {et~al.}(2020)Harris, Millman, van~der Walt, Gommers, Virtanen, Cournapeau, Wieser, Taylor, Berg, Smith, Kern, Picus, Hoyer, van Kerkwijk, Brett, Haldane, del R{\'{i}}o, Wiebe, Peterson, G{\'{e}}rard-Marchant, Sheppard, Reddy, Weckesser, Abbasi, Gohlke, \& Oliphant}]{harris2020array}
Harris, C.~R., Millman, K.~J., van~der Walt, S.~J., {et~al.} 2020, Nature, 585, 357, \dodoi{10.1038/s41586-020-2649-2}

\bibitem[{{Heller}(2012)}]{Heller_2012}
{Heller}, R. 2012, \aap, 545, L8, \dodoi{10.1051/0004-6361/201220003}

\bibitem[{{Heller} \& {Armstrong}(2014)}]{Heller_Armstrong_2014}
{Heller}, R., \& {Armstrong}, J. 2014, Astrobiology, 14, 50, \dodoi{10.1089/ast.2013.1088}

\bibitem[{{Heller} \& {Barnes}(2013)}]{Heller_Barnes_2013}
{Heller}, R., \& {Barnes}, R. 2013, Astrobiology, 13, 18, \dodoi{10.1089/ast.2012.0859}

\bibitem[{Hill {et~al.}(2023)Hill, Bott, Dalba, Fetherolf, Kane, Kopparapu, Li, \& Ostberg}]{Hill_2023}
Hill, M.~L., Bott, K., Dalba, P.~A., {et~al.} 2023, The Astronomical Journal, 165, 34, \dodoi{10.3847/1538-3881/aca1c0}

\bibitem[{{Hill} {et~al.}(2018){Hill}, {Kane}, {Seperuelo Duarte}, {Kopparapu}, {Gelino}, \& {Wittenmyer}}]{Hill_2018}
{Hill}, M.~L., {Kane}, S.~R., {Seperuelo Duarte}, E., {et~al.} 2018, \apj, 860, 67, \dodoi{10.3847/1538-4357/aac384}

\bibitem[{{Horner} {et~al.}(2020){Horner}, {Kane}, {Marshall}, {Dalba}, {Holt}, {Wood}, {Maynard-Casely}, {Wittenmyer}, {Lykawka}, {Hill}, {Salmeron}, {Bailey}, {L{\"o}hne}, {Agnew}, {Carter}, \& {Tylor}}]{hor20}
{Horner}, J., {Kane}, S.~R., {Marshall}, J.~P., {et~al.} 2020, \pasp, 132, 102001, \dodoi{10.1088/1538-3873/ab8eb9}

\bibitem[{Horvath \& Galante(2012)}]{Horvath_Galante_2012}
Horvath, J.~E., \& Galante, D. 2012, International Journal of Astrobiology, 11, 279–286, \dodoi{10.1017/S1473550412000304}

\bibitem[{Hunter(2007)}]{Hunter:2007}
Hunter, J.~D. 2007, Computing in Science \& Engineering, 9, 90, \dodoi{10.1109/MCSE.2007.55}

\bibitem[{{Ibrahim} {et~al.}(2018){Ibrahim}, {Malasan}, {Kunjaya}, {Timur Jaelani}, {Puannandra Putri}, \& {Djamal}}]{pogson_use_2018}
{Ibrahim}, I., {Malasan}, H.~L., {Kunjaya}, C., {et~al.} 2018, Research in Astronomy and Astrophysics, 18, 041, \dodoi{10.1088/1674-4527/18/4/41}

\bibitem[{{Jenkins} {et~al.}(2017){Jenkins}, {Jones}, {Tuomi}, {D{\'\i}az}, {Cordero}, {Aguayo}, {Pantoja}, {Arriagada}, {Mahu}, {Brahm}, {Rojo}, {Soto}, {Ivanyuk}, {Becerra Yoma}, {Day-Jones}, {Ruiz}, {Pavlenko}, {Barnes}, {Murgas}, {Pinfield}, {Jones}, {L{\'o}pez-Morales}, {Shectman}, {Butler}, \& {Minniti}}]{HD_165155}
{Jenkins}, J.~S., {Jones}, H.~R.~A., {Tuomi}, M., {et~al.} 2017, \mnras, 466, 443, \dodoi{10.1093/mnras/stw2811}

\bibitem[{Jim\'{e}nez-Torres {et~al.}(2013)Jim\'{e}nez-Torres, Pichardo, Lake, \& Segura}]{Torres_2013}
Jim\'{e}nez-Torres, J.~J., Pichardo, B., Lake, G., \& Segura, A. 2013, Astrobiology, 13, 491, \dodoi{10.1089/ast.2012.0842}

\bibitem[{{Kane} \& {Gelino}(2012)}]{2012PASP..124..323K}
{Kane}, S.~R., \& {Gelino}, D.~M. 2012, \pasp, 124, 323, \dodoi{10.1086/665271}

\bibitem[{Kasting {et~al.}(1993)Kasting, Whitmire, \& Reynolds}]{KASTING1993108}
Kasting, J.~F., Whitmire, D.~P., \& Reynolds, R.~T. 1993, Icarus, 101, 108, \dodoi{https://doi.org/10.1006/icar.1993.1010}

\bibitem[{Kopparapu {et~al.}(2014)Kopparapu, Ramirez, SchottelKotte, Kasting, Domagal-Goldman, \& Eymet}]{Kopparapu_2014}
Kopparapu, R.~K., Ramirez, R.~M., SchottelKotte, J., {et~al.} 2014, The Astrophysical Journal, 787, L29, \dodoi{10.1088/2041-8205/787/2/l29}

\bibitem[{Kopparapu {et~al.}(2013)Kopparapu, Ramirez, Kasting, Eymet, Robinson, Mahadevan, Terrien, Domagal-Goldman, Meadows, \& Deshpande}]{Kopparapu_2013}
Kopparapu, R.~K., Ramirez, R., Kasting, J.~F., {et~al.} 2013, The Astrophysical Journal, 765, 131, \dodoi{10.1088/0004-637X/765/2/131}

\bibitem[{Li {et~al.}(2019)Li, Mustill, \& Davies}]{10.1093/mnras/stz1794}
Li, D., Mustill, A.~J., \& Davies, M.~B. 2019, Monthly Notices of the Royal Astronomical Society, 488, 1366, \dodoi{10.1093/mnras/stz1794}

\bibitem[{Li {et~al.}(2020)Li, Mustill, \& Davies}]{10.1093/mnras/staa1622}
---. 2020, Monthly Notices of the Royal Astronomical Society, 496, 1149, \dodoi{10.1093/mnras/staa1622}

\bibitem[{{Li} {et~al.}(2011){Li}, {Leaman}, {Chornock}, {Filippenko}, {Poznanski}, {Ganeshalingam}, {Wang}, {Modjaz}, {Jha}, {Foley}, \& {Smith}}]{SNe_Rates1}
{Li}, W., {Leaman}, J., {Chornock}, R., {et~al.} 2011, \mnras, 412, 1441, \dodoi{10.1111/j.1365-2966.2011.18160.x}

\bibitem[{Lineweaver {et~al.}(2004)Lineweaver, Fenner, \& Gibson}]{lineweaver2004galactic}
Lineweaver, C.~H., Fenner, Y., \& Gibson, B.~K. 2004, Science, 303, 59

\bibitem[{Lingam \& Loeb(2017{\natexlab{a}})}]{Lingam_2017}
Lingam, M., \& Loeb, A. 2017{\natexlab{a}}, The Astrophysical Journal, 848, 41, \dodoi{10.3847/1538-4357/aa8e96}

\bibitem[{Lingam \& Loeb(2017{\natexlab{b}})}]{Lingam_2017b}
---. 2017{\natexlab{b}}, The Astrophysical Journal Letters, 846, L21, \dodoi{10.3847/2041-8213/aa8860}

\bibitem[{Lisse {et~al.}(2020)Lisse, Desch, Unterborn, Kane, Young, Hartnett, Hinkel, Shim, Mamajek, \& Izenberg}]{Lisse_2020}
Lisse, C.~M., Desch, S.~J., Unterborn, C.~T., {et~al.} 2020, The Astrophysical Journal Letters, 898, L17, \dodoi{10.3847/2041-8213/ab9b91}

\bibitem[{Loeb {et~al.}(2016)Loeb, Batista, \& Sloan}]{loeb2016relative}
Loeb, A., Batista, R.~A., \& Sloan, D. 2016, Journal of Cosmology and Astroparticle Physics, 2016, 040

\bibitem[{{Mahdi} {et~al.}(2016){Mahdi}, {Soubiran}, {Blanco-Cuaresma}, \& {Chemin}}]{solar_twins_2016}
{Mahdi}, D., {Soubiran}, C., {Blanco-Cuaresma}, S., \& {Chemin}, L. 2016, \aap, 587, A131, \dodoi{10.1051/0004-6361/201527472}

\bibitem[{{Mann} {et~al.}(2022){Mann}, {Wood}, {Schmidt}, {Barber}, {Owen}, {Tofflemire}, {Newton}, {Mamajek}, {Bush}, {Mace}, {Kraus}, {Thao}, {Vanderburg}, {Llama}, {Johns-Krull}, {Prato}, {Stahl}, {Tang}, {Fields}, {Collins}, {Collins}, {Gan}, {Jensen}, {Kamler}, {Schwarz}, {Furlan}, {Gnilka}, {Howell}, {Lester}, {Owens}, {Suarez}, {Mekarnia}, {Guillot}, {Abe}, {Triaud}, {Johnson}, {Milburn}, {Rizzuto}, {Quinn}, {Kerr}, {Ricker}, {Vanderspek}, {Latham}, {Seager}, {Winn}, {Jenkins}, {Guerrero}, {Shporer}, {Schlieder}, {McLean}, \& {Wohler}}]{TOI-1227}
{Mann}, A.~W., {Wood}, M.~L., {Schmidt}, S.~P., {et~al.} 2022, \aj, 163, 156, \dodoi{10.3847/1538-3881/ac511d}

\bibitem[{Martin \& Livio(2015)}]{Martin_2015}
Martin, R.~G., \& Livio, M. 2015, The Astrophysical Journal, 810, 105, \dodoi{10.1088/0004-637X/810/2/105}

\bibitem[{{Mayor} {et~al.}(2009){Mayor}, {Udry}, {Lovis}, {Pepe}, {Queloz}, {Benz}, {Bertaux}, {Bouchy}, {Mordasini}, \& {Segransan}}]{2009A&A...493..639M}
{Mayor}, M., {Udry}, S., {Lovis}, C., {et~al.} 2009, \aap, 493, 639, \dodoi{10.1051/0004-6361:200810451}

\bibitem[{Melott \& Thomas(2011)}]{Melott_2011}
Melott, A.~L., \& Thomas, B.~C. 2011, Astrobiology, 11, 343, \dodoi{10.1089/ast.2010.0603}

\bibitem[{{Melott} {et~al.}(2017){Melott}, {Thomas}, {Kachelrie{\ss}}, {Semikoz}, \& {Overholt}}]{Melott_2017}
{Melott}, A.~L., {Thomas}, B.~C., {Kachelrie{\ss}}, M., {Semikoz}, D.~V., \& {Overholt}, A.~C. 2017, \apj, 840, 105, \dodoi{10.3847/1538-4357/aa6c57}

\bibitem[{{Minniti} {et~al.}(2009){Minniti}, {Butler}, {L{\'o}pez-Morales}, {Shectman}, {Adams}, {Arriagada}, {Boss}, \& {Chambers}}]{HD_48265}
{Minniti}, D., {Butler}, R.~P., {L{\'o}pez-Morales}, M., {et~al.} 2009, \apj, 693, 1424, \dodoi{10.1088/0004-637X/693/2/1424}

\bibitem[{{Naoz}(2016)}]{kozai-lidov}
{Naoz}, S. 2016, \araa, 54, 441, \dodoi{10.1146/annurev-astro-081915-023315}

\bibitem[{{Narang} {et~al.}(2018){Narang}, {Manoj}, {Furlan}, {Mordasini}, {Henning}, {Mathew}, {Banyal}, \& {Sivarani}}]{nar18}
{Narang}, M., {Manoj}, P., {Furlan}, E., {et~al.} 2018, \aj, 156, 221, \dodoi{10.3847/1538-3881/aae391}

\bibitem[{{Narang} {et~al.}(2023){Narang}, {Oza}, {Hakim}, {Manoj}, {Banyal}, \& {Thorngren}}]{nar23}
{Narang}, M., {Oza}, A.~V., {Hakim}, K., {et~al.} 2023, \aj, 165, 1, \dodoi{10.3847/1538-3881/ac9eb8}

\bibitem[{{NASA Exoplanet Science Institute}(2020)}]{nea12}
{NASA Exoplanet Science Institute}. 2020, Planetary Systems Table,  IPAC, \dodoi{10.26133/NEA12}

\bibitem[{{Pecaut} \& {Mamajek}(2013)}]{Pecaut_Mamajek_2013}
{Pecaut}, M.~J., \& {Mamajek}, E.~E. 2013, \apjs, 208, 9, \dodoi{10.1088/0067-0049/208/1/9}

\bibitem[{Perkins {et~al.}(2024)Perkins, Ellis, Fields, Hartmann, Liu, McLaughlin, Surman, \& Wang}]{Perkins_2024}
Perkins, H. M.~L., Ellis, J., Fields, B.~D., {et~al.} 2024, The Astrophysical Journal, 961, 170, \dodoi{10.3847/1538-4357/ad12b7}

\bibitem[{{Perryman} {et~al.}(1997){Perryman}, {Lindegren}, {Kovalevsky}, {Hoeg}, {Bastian}, {Bernacca}, {Cr{\'e}z{\'e}}, {Donati}, {Grenon}, {Grewing}, {van Leeuwen}, {van der Marel}, {Mignard}, {Murray}, {Le Poole}, {Schrijver}, {Turon}, {Arenou}, {Froeschl{\'e}}, \& {Petersen}}]{Hipparcos1997}
{Perryman}, M.~A.~C., {Lindegren}, L., {Kovalevsky}, J., {et~al.} 1997, \aap, 323, L49

\bibitem[{{Pogson}(1856)}]{1856MNRAS..17...12P}
{Pogson}, N. 1856, \mnras, 17, 12, \dodoi{10.1093/mnras/17.1.12}

\bibitem[{{Portegies Zwart} {et~al.}(2021){Portegies Zwart}, {Torres}, {Cai}, \& {Brown}}]{Oort_Cloud_Zwart}
{Portegies Zwart}, S., {Torres}, S., {Cai}, M.~X., \& {Brown}, A. G.~A. 2021, \aap, 652, A144, \dodoi{10.1051/0004-6361/202040096}

\bibitem[{Ramachandran \& Tsokos(2021)}]{Bootstrapping}
Ramachandran, K.~M., \& Tsokos, C.~P. 2021, in Mathematical Statistics with Applications in R (Third Edition), third edition edn., ed. K.~M. Ramachandran \& C.~P. Tsokos (Academic Press), 531--568, \dodoi{https://doi.org/10.1016/B978-0-12-817815-7.00013-0}

\bibitem[{{Ram{\'\i}rez} {et~al.}(2009){Ram{\'\i}rez}, {Mel{\'e}ndez}, \& {Asplund}}]{solar_twins_2009}
{Ram{\'\i}rez}, I., {Mel{\'e}ndez}, J., \& {Asplund}, M. 2009, \aap, 508, L17, \dodoi{10.1051/0004-6361/200913038}

\bibitem[{{Raymond} {et~al.}(2023){Raymond}, {Izidoro}, \& {Kaib}}]{Oort_Exoplanets}
{Raymond}, S.~N., {Izidoro}, A., \& {Kaib}, N.~A. 2023, \mnras, 524, L72, \dodoi{10.1093/mnrasl/slad079}

\bibitem[{{Recio-Blanco} {et~al.}(2023){Recio-Blanco}, {de Laverny}, {Palicio}, {Kordopatis}, {{\'A}lvarez}, {Schultheis}, {Contursi}, {Zhao}, {Torralba Elipe}, {Ordenovic}, {Manteiga}, {Dafonte}, {Oreshina-Slezak}, {Bijaoui}, {Fr{\'e}mat}, {Seabroke}, {Pailler}, {Spitoni}, {Poggio}, {Creevey}, {Abreu Aramburu}, {Accart}, {Andrae}, {Bailer-Jones}, {Bellas-Velidis}, {Brouillet}, {Brugaletta}, {Burlacu}, {Carballo}, {Casamiquela}, {Chiavassa}, {Cooper}, {Dapergolas}, {Delchambre}, {Dharmawardena}, {Drimmel}, {Edvardsson}, {Fouesneau}, {Garabato}, {Garc{\'\i}a-Lario}, {Garc{\'\i}a-Torres}, {Gavel}, {Gomez}, {Gonz{\'a}lez-Santamar{\'\i}a}, {Hatzidimitriou}, {Heiter}, {Jean-Antoine Piccolo}, {Kontizas}, {Korn}, {Lanzafame}, {Lebreton}, {Le Fustec}, {Licata}, {Lindstr{\o}m}, {Livanou}, {Lobel}, {Lorca}, {Magdaleno Romeo}, {Marocco}, {Marshall}, {Mary}, {Nicolas}, {Pallas-Quintela}, {Panem}, {Pichon}, {Riclet}, {Robin}, {Rybizki}, {Santove{\~n}a}, {Silvelo}, {Smart}, {Sarro}, {Sordo}, {Soubiran}, {S{\"u}veges},
  {Ulla}, {Vallenari}, {Zorec}, {Utrilla}, \& {Bakker}}]{bla23}
{Recio-Blanco}, A., {de Laverny}, P., {Palicio}, P.~A., {et~al.} 2023, \aap, 674, A29, \dodoi{10.1051/0004-6361/202243750}

\bibitem[{{Redfield} {et~al.}(2024){Redfield}, {Batalha}, {Benneke}, {Biller}, {Espinoza}, {France}, {Konopacky}, {Kreidberg}, {Rauscher}, \& {Sing}}]{red2024}
{Redfield}, S., {Batalha}, N., {Benneke}, B., {et~al.} 2024, arXiv e-prints, arXiv:2404.02932, \dodoi{10.48550/arXiv.2404.02932}

\bibitem[{Reyl{\'e} {et~al.}(2021)Reyl{\'e}, Jardine, Fouqu{\'e}, Caballero, Smart, \& Sozzetti}]{reyle202110}
Reyl{\'e}, C., Jardine, K., Fouqu{\'e}, P., {et~al.} 2021, Astronomy \& Astrophysics, 650, A201, \dodoi{10.1051/0004-6361/202140985}

\bibitem[{{Reyl{\'e}} {et~al.}(2021){Reyl{\'e}}, {Jardine}, {Fouqu{\'e}}, {Caballero}, {Smart}, \& {Sozzetti}}]{gaia_10pc}
{Reyl{\'e}}, C., {Jardine}, K., {Fouqu{\'e}}, P., {et~al.} 2021, \aap, 650, A201, \dodoi{10.1051/0004-6361/202140985}

\bibitem[{{Reyl{\'e}} {et~al.}(2022){Reyl{\'e}}, {Jardine}, {Fouqu{\'e}}, {Caballero}, {Smart}, \& {Sozzetti}}]{Gaia_10pc_Update}
{Reyl{\'e}}, C., {Jardine}, K., {Fouqu{\'e}}, P., {et~al.} 2022, in The 21st Cambridge Workshop on Cool Stars, Stellar Systems, and the Sun, Cambridge Workshop on Cool Stars, Stellar Systems, and the Sun, 218, \dodoi{10.5281/zenodo.7669746}

\bibitem[{{Rickman} {et~al.}(2023){Rickman}, {Wajer}, {Przy{\l}uski}, {Wi{\'s}niowski}, {Nesvorn{\'y}}, \& {Morbidelli}}]{rickman_2023}
{Rickman}, H., {Wajer}, P., {Przy{\l}uski}, R., {et~al.} 2023, \mnras, 520, 637, \dodoi{10.1093/mnras/stac3705}

\bibitem[{Rickman {et~al.}(2022)Rickman, Wajer, Przyłuski, Wiśniowski, Nesvorný, \& Morbidelli}]{10.1093/mnras/stac3705}
Rickman, H., Wajer, P., Przyłuski, R., {et~al.} 2022, Monthly Notices of the Royal Astronomical Society, 520, 637, \dodoi{10.1093/mnras/stac3705}

\bibitem[{{Rodr{\'\i}guez-Mozos} \& {Moya}(2019)}]{magnetospheres}
{Rodr{\'\i}guez-Mozos}, J.~M., \& {Moya}, A. 2019, \aap, 630, A52, \dodoi{10.1051/0004-6361/201935543}

\bibitem[{{Rybizki} {et~al.}(2022){Rybizki}, {Green}, {Rix}, {El-Badry}, {Demleitner}, {Zari}, {Udalski}, {Smart}, \& {Gould}}]{Rybizki_2022}
{Rybizki}, J., {Green}, G.~M., {Rix}, H.-W., {et~al.} 2022, \mnras, 510, 2597, \dodoi{10.1093/mnras/stab3588}

\bibitem[{{Saydjari} {et~al.}(2023){Saydjari}, {Schlafly}, {Lang}, {Meisner}, {Green}, {Zucker}, {Zelko}, {Speagle}, {Daylan}, {Lee}, {Valdes}, {Schlegel}, \& {Finkbeiner}}]{decaps}
{Saydjari}, A.~K., {Schlafly}, E.~F., {Lang}, D., {et~al.} 2023, \apjs, 264, 28, \dodoi{10.3847/1538-4365/aca594}

\bibitem[{{Schneider} {et~al.}(2011){Schneider}, {Dedieu}, {Le Sidaner}, {Savalle}, \& {Zolotukhin}}]{Schneider_2011}
{Schneider}, J., {Dedieu}, C., {Le Sidaner}, P., {Savalle}, R., \& {Zolotukhin}, I. 2011, \aap, 532, A79, \dodoi{10.1051/0004-6361/201116713}

\bibitem[{Schulze-Makuch {et~al.}(2011)Schulze-Makuch, M\'{e}ndez, Fair\'{e}n, von Paris, Turse, Boyer, Davila, Ant\'{o}nio, Catling, \& Irwin}]{Schulze_2011}
Schulze-Makuch, D., M\'{e}ndez, A., Fair\'{e}n, A.~G., {et~al.} 2011, Astrobiology, 11, 1041, \dodoi{10.1089/ast.2010.0592}

\bibitem[{Schwieterman {et~al.}(2018)Schwieterman, Kiang, Parenteau, Harman, DasSarma, Fisher, Arney, Hartnett, Reinhard, Olson, Meadows, Cockell, Walker, Grenfell, Hegde, Rugheimer, Hu, \& Lyons}]{doi:10.1089/ast.2017.1729}
Schwieterman, E.~W., Kiang, N.~Y., Parenteau, M.~N., {et~al.} 2018, Astrobiology, 18, 663, \dodoi{10.1089/ast.2017.1729}

\bibitem[{Spinelli \& Ghirlanda(2023)}]{Spinelli_2023}
Spinelli, R., \& Ghirlanda, G. 2023, Universe, 9, \dodoi{10.3390/universe9020060}

\bibitem[{Spinelli {et~al.}(2021)Spinelli, Ghirlanda, Haardt, Ghisellini, \& Scuderi}]{spinelli2021best}
Spinelli, R., Ghirlanda, G., Haardt, F., Ghisellini, G., \& Scuderi, G. 2021, Astronomy \& Astrophysics, 647, A41

\bibitem[{Spitoni {et~al.}(2017)Spitoni, Gioannini, \& Matteucci}]{spitoni2017galactic}
Spitoni, E., Gioannini, L., \& Matteucci, F. 2017, Astronomy \& Astrophysics, 605, A38

\bibitem[{Stark {et~al.}(2014)Stark, Roberge, Mandell, \& Robinson}]{sta2014}
Stark, C.~C., Roberge, A., Mandell, A., \& Robinson, T.~D. 2014, The Astrophysical Journal, 795, 122, \dodoi{10.1088/0004-637X/795/2/122}

\bibitem[{{Swastik} {et~al.}(2022){Swastik}, {Banyal}, {Narang}, {Manoj}, {Sivarani}, {Rajaguru}, {Unni}, \& {Banerjee}}]{swa2023}
{Swastik}, C., {Banyal}, R.~K., {Narang}, M., {et~al.} 2022, \aj, 164, 60, \dodoi{10.3847/1538-3881/ac756a}

\bibitem[{{Swastik} {et~al.}(2021){Swastik}, {Banyal}, {Narang}, {Manoj}, {Sivarani}, {Reddy}, \& {Rajaguru}}]{2021AJ....161..114S}
---. 2021, \aj, 161, 114, \dodoi{10.3847/1538-3881/abd802}

\bibitem[{Swastik {et~al.}(2023)Swastik, Banyal, Narang, Unni, Banerjee, Manoj, \& Sivarani}]{swastik2023age}
Swastik, C., Banyal, R.~K., Narang, M., {et~al.} 2023, The Astronomical Journal, 166, 91

\bibitem[{{Swastik} {et~al.}(2024){Swastik}, {Banyal}, {Narang}, {Unni}, \& {Sivarani}}]{2024AJ....167..270S}
{Swastik}, C., {Banyal}, R.~K., {Narang}, M., {Unni}, A., \& {Sivarani}, T. 2024, \aj, 167, 270, \dodoi{10.3847/1538-3881/ad40ae}

\bibitem[{Thomas {et~al.}(2005{\natexlab{a}})Thomas, Jackman, Melott, Laird, Stolarski, Gehrels, Cannizzo, \& Hogan}]{Thomas_2005}
Thomas, B.~C., Jackman, C.~H., Melott, A.~L., {et~al.} 2005{\natexlab{a}}, The Astrophysical Journal, 622, L153, \dodoi{10.1086/429799}

\bibitem[{{Thomas} \& {Melott}(2006)}]{Thomas_Melott_2006}
{Thomas}, B.~C., \& {Melott}, A.~L. 2006, New Journal of Physics, 8, 120, \dodoi{10.1088/1367-2630/8/7/120}

\bibitem[{{Thomas} \& {Yelland}(2023)}]{SNe_Updated_2023}
{Thomas}, B.~C., \& {Yelland}, A.~M. 2023, \apj, 950, 41, \dodoi{10.3847/1538-4357/accf8a}

\bibitem[{Thomas {et~al.}(2005{\natexlab{b}})Thomas, Melott, Jackman, Laird, Medvedev, Stolarski, Gehrels, Cannizzo, Hogan, \& Ejzak}]{Thomas_2005a}
Thomas, B.~C., Melott, A.~L., Jackman, C.~H., {et~al.} 2005{\natexlab{b}}, The Astrophysical Journal, 634, 509, \dodoi{10.1086/496914}

\bibitem[{{Tinetti} {et~al.}(2018){Tinetti}, {Drossart}, {Eccleston}, {Hartogh}, {Heske}, {Leconte}, {Micela}, {Ollivier}, {Pilbratt}, {Puig}, {Turrini}, {Vandenbussche}, {Wolkenberg}, {Beaulieu}, {Buchave}, {Ferus}, {Griffin}, {Guedel}, {Justtanont}, {Lagage}, {Machado}, {Malaguti}, {Min}, {N{\o}rgaard-Nielsen}, {Rataj}, {Ray}, {Ribas}, {Swain}, {Szabo}, {Werner}, {Barstow}, {Burleigh}, {Cho}, {Coud{\'e} du Foresto}, {Coustenis}, {Decin}, {Encrenaz}, {Galand}, {Gillon}, {Helled}, {Morales}, {Garc{\'\i}a Mu{\~n}oz}, {Moneti}, {Pagano}, {Pascale}, {Piccioni}, {Pinfield}, {Sarkar}, {Selsis}, {Tennyson}, {Triaud}, {Venot}, {Waldmann}, {Waltham}, {Wright}, {Amiaux}, {Augu{\`e}res}, {Berth{\'e}}, {Bezawada}, {Bishop}, {Bowles}, {Coffey}, {Colom{\'e}}, {Crook}, {Crouzet}, {Da Peppo}, {Sanz}, {Focardi}, {Frericks}, {Hunt}, {Kohley}, {Middleton}, {Morgante}, {Ottensamer}, {Pace}, {Pearson}, {Stamper}, {Symonds}, {Rengel}, {Renotte}, {Ade}, {Affer}, {Alard}, {Allard}, {Altieri}, {Andr{\'e}}, {Arena}, {Argyriou},
  {Aylward}, {Baccani}, {Bakos}, {Banaszkiewicz}, {Barlow}, {Batista}, {Bellucci}, {Benatti}, {Bernardi}, {B{\'e}zard}, {Blecka}, {Bolmont}, {Bonfond}, {Bonito}, {Bonomo}, {Brucato}, {Brun}, {Bryson}, {Bujwan}, {Casewell}, {Charnay}, {Pestellini}, {Chen}, {Ciaravella}, {Claudi}, {Cl{\'e}dassou}, {Damasso}, {Damiano}, {Danielski}, {Deroo}, {Di Giorgio}, {Dominik}, {Doublier}, {Doyle}, {Doyon}, {Drummond}, {Duong}, {Eales}, {Edwards}, {Farina}, {Flaccomio}, {Fletcher}, {Forget}, {Fossey}, {Fr{\"a}nz}, {Fujii}, {Garc{\'\i}a-Piquer}, {Gear}, {Geoffray}, {G{\'e}rard}, {Gesa}, {Gomez}, {Graczyk}, {Griffith}, {Grodent}, {Guarcello}, {Gustin}, {Hamano}, {Hargrave}, {Hello}, {Heng}, {Herrero}, {Hornstrup}, {Hubert}, {Ida}, {Ikoma}, {Iro}, {Irwin}, {Jarchow}, {Jaubert}, {Jones}, {Julien}, {Kameda}, {Kerschbaum}, {Kervella}, {Koskinen}, {Krijger}, {Krupp}, {Lafarga}, {Landini}, {Lellouch}, {Leto}, {Luntzer}, {Rank-L{\"u}ftinger}, {Maggio}, {Maldonado}, {Maillard}, {Mall}, {Marquette}, {Mathis}, {Maxted}, {Matsuo},
  {Medvedev}, {Miguel}, {Minier}, {Morello}, {Mura}, {Narita}, {Nascimbeni}, {Nguyen Tong}, {Noce}, {Oliva}, {Palle}, {Palmer}, {Pancrazzi}, {Papageorgiou}, {Parmentier}, {Perger}, {Petralia}, {Pezzuto}, {Pierrehumbert}, {Pillitteri}, {Piotto}, {Pisano}, {Prisinzano}, {Radioti}, {R{\'e}ess}, {Rezac}, {Rocchetto}, {Rosich}, {Sanna}, {Santerne}, {Savini}, {Scandariato}, {Sicardy}, {Sierra}, {Sindoni}, {Skup}, {Snellen}, {Sobiecki}, {Soret}, {Sozzetti}, {Stiepen}, {Strugarek}, {Taylor}, {Taylor}, {Terenzi}, {Tessenyi}, {Tsiaras}, {Tucker}, {Valencia}, {Vasisht}, {Vazan}, {Vilardell}, {Vinatier}, {Viti}, {Waters}, {Wawer}, {Wawrzaszek}, {Whitworth}, {Yung}, {Yurchenko}, {Zapatero Osorio}, {Zellem}, {Zingales}, \& {Zwart}}]{tin2018}
{Tinetti}, G., {Drossart}, P., {Eccleston}, P., {et~al.} 2018, Experimental Astronomy, 46, 135, \dodoi{10.1007/s10686-018-9598-x}

\bibitem[{{Torra} {et~al.}(2021){Torra}, {Casta{\~n}eda}, {Fabricius}, {Lindegren}, {Clotet}, {Gonz{\'a}lez-Vidal}, {Bartolom{\'e}}, {Bastian}, {Bernet}, {Biermann}, {Garralda}, {G{\'u}rpide}, {Lammers}, {Portell}, \& {Torra}}]{F_Torra_2021}
{Torra}, F., {Casta{\~n}eda}, J., {Fabricius}, C., {et~al.} 2021, \aap, 649, A10, \dodoi{10.1051/0004-6361/202039637}

\bibitem[{{Tuomi} {et~al.}(2013){Tuomi}, {Anglada-Escud{\'e}}, {Gerlach}, {Jones}, {Reiners}, {Rivera}, {Vogt}, \& {Butler}}]{2013A&A...549A..48T}
{Tuomi}, M., {Anglada-Escud{\'e}}, G., {Gerlach}, E., {et~al.} 2013, \aap, 549, A48, \dodoi{10.1051/0004-6361/201220268}

\bibitem[{{Underwood} {et~al.}(2003){Underwood}, {Jones}, \& {Sleep}}]{under2003}
{Underwood}, D.~R., {Jones}, B.~W., \& {Sleep}, P.~N. 2003, International Journal of Astrobiology, 2, 289, \dodoi{10.1017/S1473550404001715}

\bibitem[{{Unni} {et~al.}(2022){Unni}, {Narang}, {Sivarani}, {Manoj}, {Banyal}, {Surya}, {Rajaguru}, \& {Swastik}}]{uni22}
{Unni}, A., {Narang}, M., {Sivarani}, T., {et~al.} 2022, \aj, 164, 181, \dodoi{10.3847/1538-3881/ac8b7c}

\bibitem[{{van Leeuwen}(2007)}]{Hipparcos_2007}
{van Leeuwen}, F. 2007, \aap, 474, 653, \dodoi{10.1051/0004-6361:20078357}

\bibitem[{{Wang} {et~al.}(2020){Wang}, {Perna}, \& {Leigh}}]{Wang_2020}
{Wang}, Y.-H., {Perna}, R., \& {Leigh}, N. W.~C. 2020, \mnras, 496, 1453, \dodoi{10.1093/mnras/staa1627}

\bibitem[{{Wang, Jifei} \& {Zhong, Zehao}(2018)}]{mass_lum}
{Wang, Jifei}, \& {Zhong, Zehao}. 2018, A\&A, 619, L1, \dodoi{10.1051/0004-6361/201834109}

\bibitem[{Ware {et~al.}(2022)Ware, Young, Truitt, \& Spacek}]{Ware_2022}
Ware, A., Young, P., Truitt, A., \& Spacek, A. 2022, The Astrophysical Journal, 929, 143, \dodoi{10.3847/1538-4357/ac5c4e}

\bibitem[{{Winn} \& {Fabrycky}(2015)}]{win15}
{Winn}, J.~N., \& {Fabrycky}, D.~C. 2015, \araa, 53, 409, \dodoi{10.1146/annurev-astro-082214-122246}

\bibitem[{{Zhu} \& {Dong}(2021)}]{zhu21}
{Zhu}, W., \& {Dong}, S. 2021, \araa, 59, 291, \dodoi{10.1146/annurev-astro-112420-020055}

\end{thebibliography}

\appendix

\section{Function to obtain 25~pc samples}\label{10pc_code}
\begin{verbatim}
    from astroquery.gaia import Gaia as ga

    def nearby25(par,dec,ra):
        query = "SELECT * \
                FROM gaiadr3.gaia_source as g\
                JOIN gaiadr3.astrophysical_parameters AS ap ON g.source_id = ap.source_id\
                WHERE SQRT(POWER((1000/g.parallax)*COS(RADIANS(g.dec))*COS(RADIANS(g.ra)) 
                - 1000/"+str(par)+"*COS(RADIANS("+str(dec)+"))*COS(RADIANS("+str(ra)+")),2) +
                POWER((1000/g.parallax)*COS(RADIANS(g.dec))*SIN(RADIANS(g.ra)) 
                - 1000/"+str(par)+"*COS(RADIANS("+str(dec)+"))*SIN(RADIANS("+str(ra)+")),2) + 
                POWER((1000/g.parallax)*SIN(RADIANS(g.dec))
                - 1000/"+str(par)+"*SIN(RADIANS("+str(dec)+")),2)) <= 25 AND g.parallax > 0"
        
        r = ga.launch_job_async(query)
        res = r.get_results()
        
        return res
\end{verbatim}

The above code defines a function {\tt{nearby25}} that queries the {\it Gaia DR3} archive to find stars within a 25~pc radius of a habitable zone system based on its {\tt{parallax}} (mas), {\tt{ra}} and {\tt{dec}}. The {\it Gaia} parallax ($\varpi$) is converted to the radial distance from the Sun by simple inversion $r = 1/\varpi$. The query also joins data from two \textit{GAIA DR3} tables: {\tt{gaia\_source}} and {\tt{astrophysical\_parameters}}, based on a common {\tt{source\_id}}.

The code converts the spherical polar coordinates $(r,\theta,\phi)$ to Cartesian coordinates $(x,y,z)$. It makes the following assumptions to do so:
\begin{itemize}
  \item The $x$-axis extends from the origin $O$ through the vernal equinox along the equatorial plane. This axis aligns with 0$^\circ$ RA.
  \item The $y$-axis, orthogonal to the $x$-axis within the equatorial plane, points towards 90$^\circ$ RA.
  \item The $z$-axis, is perpendicular to the equatorial plane.
\end{itemize}

To be consistency with spherical geometry, we  define, $\phi = \textrm{RA}$ (angle made with x-axis) and $\theta = 90 - \textrm{Dec}$ (angle made with z-axis). These spherical polar coordinates $(r,\theta,\phi)$ are then converted to Cartesian coordinates $(x,y,z)$ via the following transformation:

\begin{equation}
\begin{aligned}
    x &= r \sin{\theta}\cos{\phi} \\
    y &= r \sin{\theta}\sin{\phi} \\
    z &= r \cos{\theta}
\end{aligned}
\end{equation}

The distance between the HZ star $(x,y,z)$ and a neighboring star $(x',y',z')$ in 10~pc region is calculated using: 
$d=\sqrt{(x - x')^2 + (y - y')^2 + (z - z')^2}$. In summary, for each HZ system, the three input parameters -- parallax, declination and right ascension define its spatial location. \textit{Gaia DR3} is queried for the spatial locations of stars detected within 25~pc of the HZS using the {\tt{nearby25}} function which is defined above. This 25~pc volume of stars around a HZ system is then used to construct the 10~pc neighborhood of stars using the bootstrapping algorithm as explained in the Appendix \ref{derive}.

\section{Constructing the 10~pc stellar neighborhood using bootstrap method }\label{derive}
    \begin{figure}
        \centering
        \includegraphics[width =0.75\columnwidth]{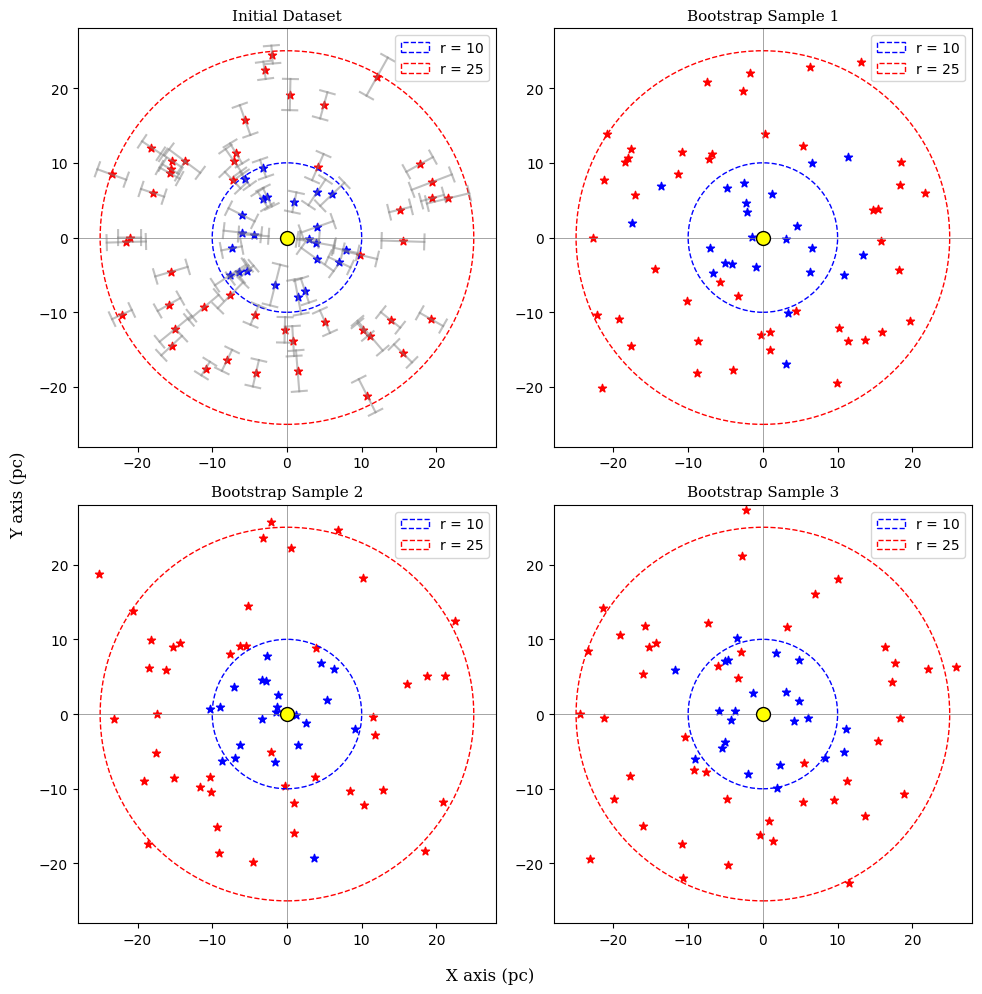}
        \caption{Top left: The initial dataset of stars with measured parallaxes $(\varpi)$  and $\pm 1\sigma_{\varpi}$  errors projected on a 2D plane. The blue symbols are stars residing within a 10~pc radius denoted by blue circle and red symbols are stars in 25~pc radius denoted by red circle. Top right panel and bottom row: Three randomly drawn samples from the original distribution showing some of the stars crossing the 10~pc boundary.}
        \label{bootstrp_dem}
    \end{figure}

\begin{figure}[htpb]
\centering
    
\begin{tikzpicture}[node distance=2.1cm, auto][b!]
\node (read2) [process,text width=6cm] {{\it Gaia} data of $n$ stars in the 25~pc neighborhood of HZS};
\node (generate) [process, below of=read2, text width=6cm] {Draw random sample of $n$ stars from Normal distribution \[
\left\{ N(\varpi_j, \sigma_{\varpi _j}) \right\}_{j=1}^n
\]};
\node (subset) [process, below of=generate,text width=6cm] {Select \(m\) stars with \(d \leq 10\) to construct a 10~pc neighborhood};
\node (analysis) [process, below of=subset, text width=6cm] {Compute and store astrophysical parameters for $m$ stars};
\node (loop) [decision, below of=analysis] {i $<$ 100,000};
\node (store) [process, below of=loop, text width=6cm] {Analyze the sampling distribution of astrophysical parameters of 10~pc regions around HZS};

\draw [arrow] (read2) -- (generate);
\draw [arrow] (generate) -- (subset);
\draw [arrow] (subset) -- (analysis);
\draw [arrow] (analysis) -- (loop);
\draw [arrow] (loop) -- node[anchor=east] {NO} (store);
\draw [arrow] (loop.west) -- ++(-2.7,0) |- node[pos=0.01, anchor=west] {YES} (read2);
\end{tikzpicture}

\caption{Schematic of the bootstrapping algorithm used for constructing a 10~pc neighborhood of HZ planet-hosting stars.}
    \label{flow_chart}
\end{figure}
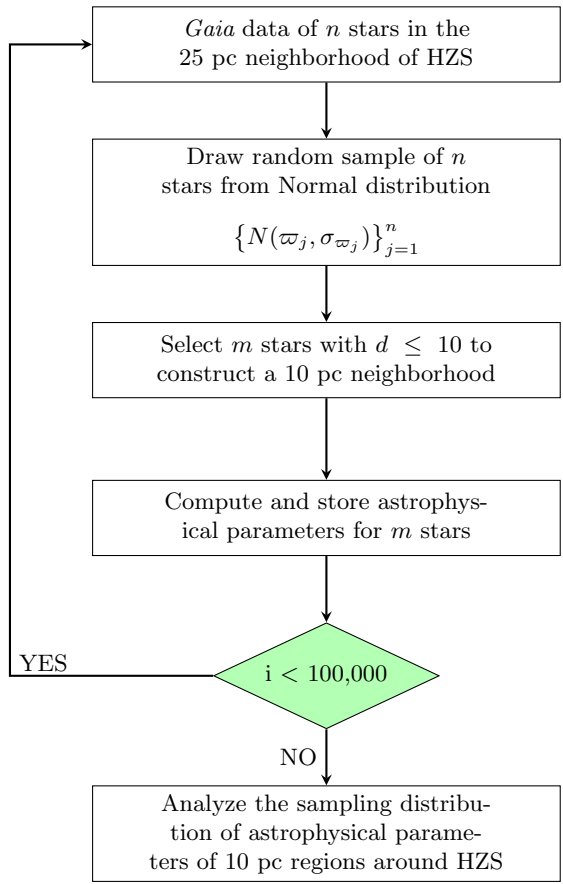

Bootstrapping is a resampling technique employed in statistics to estimate the sampling distribution of a quantity by creating multiple simulated datasets from an original sample. This process involves randomly selecting data points 
from the original dataset with replacement, constructing new samples of the same size as the original \citep{Bootstrap_Recent_Devs,Bootstrapping}.
To construct the 10~pc neighborhood of each HZS, we applied the bootstrap method to {\it Gaia}-detected stars within their 25~pc volume. This algorithm is shown as a flowchart in Figure~\ref{flow_chart}. For visualizing how the algorithm resamples to construct new configurations of neighborhoods in each iteration, we consider a hypothetical dataset of $n$ neighborhood stars (original 25~pc data from {\it Gaia}) with known parallaxes and parallax errors  as shown in the upper left panel of Figure~\ref{bootstrp_dem}. The blue and red circles represent 
the 10~pc and the 25~pc neighborhood boundary, respectively. We assume the measured parallax of each star $j$ follows the Gaussian distribution $\mathcal{N}_j (\varpi_j, \sigma_{\varpi j})$, where  $\varpi_j$ is the mean parallax and $\sigma_{\varpi j}$ is the standard deviation. In bootstrap, we randomly draw $n$ stars from the original sample $\left \{\mathcal{N}_j (\varpi_j, \sigma_{\varpi j} \right \}_{j=1}^{n}$ and select those for which $r = 1/\varpi \leq 10$~pc. 
Applying this cutoff cause some stars originally outside the 10~pc boundary to move inside, while some stars initially within the boundary to move beyond the 10~pc limit. This creates a new sample of 10~pc neighborhood. Three random realizations of bootstrap process are illustrated in top-right panel and the bottom row of Figure~\ref{bootstrp_dem}. The likelihood of a star crossing the 10 pc boundary strongly depends on the original parallax $\varpi$ and the associated parallax error $\sigma_{\varpi}$. This means that, for each random draw, stars near the 10~pc boundary in the original sample are more likely to move in and out of the 10~pc region compared to stars significantly closer to or farther from the center. For each iteration we count the number of stars in 10~pc region and also keep track of their astrophysical parameters. After 10$^5$ bootstrap runs we get a sampling distribution of each parameter. The median and standard deviation of these distributions are taken as parameters of interest (e.g., dispersion velocity, number density, etc). 

\begin{figure}[htpb]
    \centering
    \includegraphics[width=0.75\columnwidth]{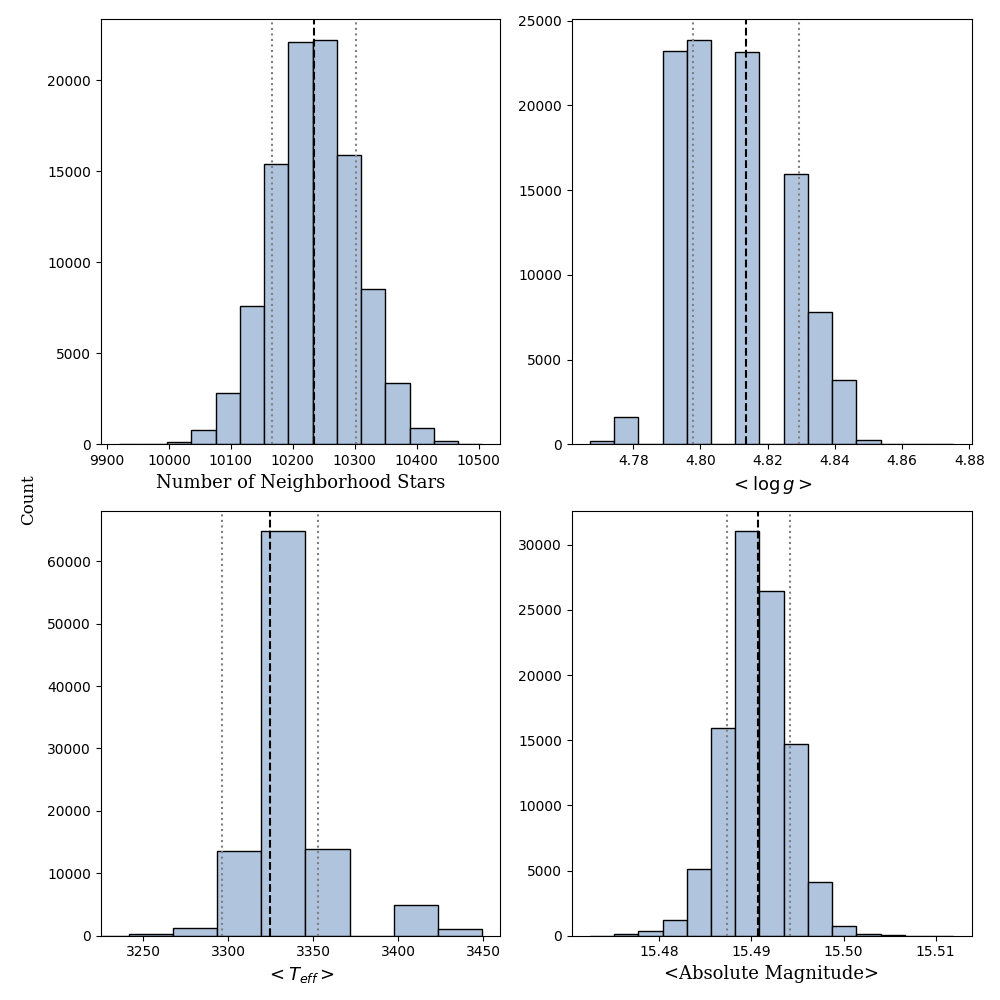}
    \caption{Bootstrapped sampling distributions of neighborhood star count (top-left panel) and astrophysical parameters: $\langle \log g \rangle$ (top-right panel), $\langle\textrm{T}_\textrm{eff}\rangle$ (bottom-left panel) and $\langle \textrm{Absolute Magnitude}\rangle$ (bottom-right panel)  of the 10~pc environment of HD~165155. Gray dotted lines represent the standard deviation and the black dashed lines represent the median.}
    \label{fig:bootstrap_example}
\end{figure}

This is a quantitative way to statistically estimate the uncertainties introduced by parallax errors in constructing the 10~pc neighborhood of HZS. The bootstrapping algorithm described above provides us with the sampling distribution of neighborhood astrophysical parameters used for our analysis. For example, the results for a system HD 165155 have been shown in Figure~\ref{fig:bootstrap_example}. For more than 20000 realizations, the neighborhood is found to have ~10200 stars, while on none of the occasions do we find 10500 or more stars. The median and standard deviation have been obtained from these sampling distributions for star count and astrophysical parameters such as $\langle \log g \rangle$, $\langle \textrm{T}_\textrm{eff} \rangle$, and $\langle \textrm{Absolute Magnitude} \rangle$. We conclude from this analysis that HD~165155 has $10235\pm 67$ stars with a median $T_{\textrm{eff}}$ of $3324\pm 28$~K, median log g of $4.81\pm 0.01$ and median absolute magnitude of $15.490 \pm 0.003$. The astrophysical parameters of all 84~HZS and their 10~pc neighborhood stars, estimated using bootstrap, are provided in the machine-readable table.

\section{10~pc Neighborhoods with High Densities}\label{high-dense-nbh}

We have studied the 10~pc neighborhood of 84~HZS and identified three systems--HD~165155, HD~159868, and HD~188015--that stand out due to their unusually high star densities. HD~165155, located $\sim$63 pc away, with $\sim$10,000 stars in its 10~pc vicinity, has a local neighborhood star density 30 times larger than the Sun and the majority of other HZS studied. HD~159868 and HD~188015, while showing lower densities, still contain $\sim$3,400 and $\sim$2,700 stars, respectively, which is about 10 times greater than the local density of the Sun's neighborhood. Given their positions at or near the galactic plane, these high-density regions might initially seem unrealistic, raising the possibility of inclusion of spurious sources or contamination of the sample by the background stars. Such artifacts can also arise from diffraction spikes of bright stars, source confusion, faults in telescope behavior, or transiting solar system objects \citep{F_Torra_2021}. We  carefully analyzed these high-density regions to address the serious concern of sample contamination.

We first confirmed that all three high-density HZS are indeed located near the galactic plane (see Figure~\ref{POS}), where the background star density is particularly high. In addition to HD~165155, HD~159868, and HD~188015, there are five other HZS located in the foreground of the high-density region of the galactic plane, though their 10~pc volume does not differ significantly from that of the Sun. This dissimilarity implies that the neighborhood data compiled from Gaia DR3 is highly accurate and reliable, even for HZS located near the galactic plane.

We also find that the distance between HD~165155 and HD~159868 is 16.5~pc, indicating that both systems are embedded in a similar stellar environment. In contrast, HD~188015 is about 60~pc away from these systems and roughly 50~pc from the Sun, which happens to be a relatively less dense environment. 

\begin{figure}[htpb]
    \centering
    \includegraphics[width=0.65\columnwidth]{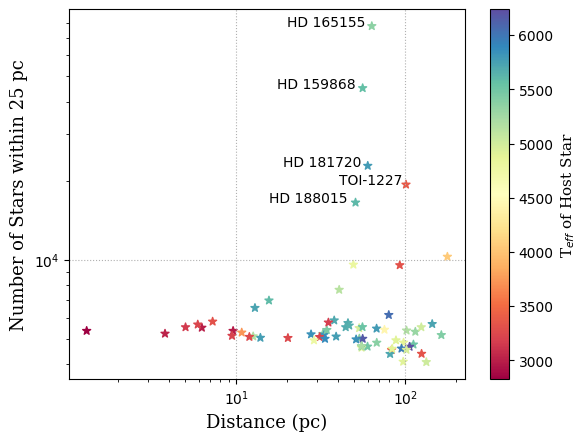}
    \caption{The 25~pc neighborhood star count for 84 HZS and their distance from Sun. The colorbar represents the effective temperature T$_\textrm{eff}$ (K) of HZ stars. The names of 5~HZS with highest star count are also labeled.}
    \label{25pc_nbh}
\end{figure}

\cite{Rybizki_2022}, developed a classifier for Gaia eDR3 sources to identify astrometric solutions as either \texttt{good} or \texttt{bad}. They set the \texttt{parallax\_over\_error} threshold $>$4.5 for high signal-to-noise ratio regimes. In our data $\sim$99\% of the sources within the 10~pc neighborhood of these high-density regions fall in that regime. Moreover, their analysis indicates that at a threshold of \texttt{parallax\_over\_error} $>$10, the number of \texttt{bad} sources is $<$10\%. This serves as a strong constraint for good astrometric sources.

\begin{figure}[htpb]
    \centering
    \includegraphics[width=0.6\columnwidth]{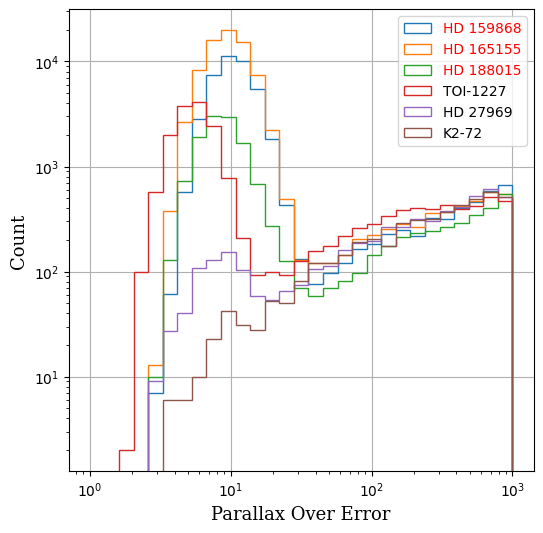}
    \caption{Distribution of \texttt{parallax\_over\_error} for stars in 25 pc neighborhood around 6 HZS. The names of 3 HZS with the highest star count in their 10~pc regions are indicated in the legend with red.}
    \label{parallax_over_error}
\end{figure}

As an additional and independent test, we also employed  Monte Carlo simulations to estimate the statistical likelihood of observing high stellar density regions within the 100~pc solar neighborhood. To this end, a raw sample of $\sim$~570,000 stars was compiled by selecting all stellar sources with a Gaia-measured parallax of $\geq$10~mas.  From this 100~pc sample, we randomly selected points in RA [0,360], Dec [-90,90], and parallax [11.11,1000] mas, and for each point, we counted the number of stars within its 10~pc neighborhood. This process was repeated one million times. We found that the probability of obtaining $>$10,000 stars in 10~pc volume around random locations is 0.14\% and that of obtaining $>$2000 is 1.7\%. These results suggest that while the neighborhoods of the three HZS in question seem rare, occurrence of dense regions of stars within 100~pc of Sun is not entirely improbable.

Finally, we plot the 25~pc neighborhood  star count for the 84~HZS in Figure~\ref{25pc_nbh}. Similar to the 10~pc region, the 25~pc  star count around HD~165155 remains highest ($\sim$77,000), followed by $\sim$45,000 stars around HD~159868. However, the 15-fold increase in volume resulted in less than an 8-fold increase in star count, indicating a notable density drop in going from 10~pc to 25~pc region. We would not expect such a drop if the background stars near the galactic plane were contaminating the neighborhood sample. 

Figure \ref{parallax_over_error} represents the distribution of the \texttt{parallax\_over\_error} for sources within the 25~pc radius of 3 HZS that have the highest population within their 10~pc radius and 3 HZS that have a lower population ($<$1000) in their 10~pc neighborhood. This plot indicates that most environments have a majority of sources with high signal to noise ratio (\texttt{parallax\_over\_error}$\gtrsim$10) and hence have good astrometric solutions \citep{Rybizki_2022}. The sharp decline on the left slope also suggests the presence of a smaller number of high parallax error sources. A value $\geq$10 indicates that the parallax error does not result in a distance error $\geq$10~pc for any source within 100~pc. The closer the object, the lesser is the error in distance. In conclusion, although our neighborhood sample may not be entirely free from spurious sources, the robust astrometric solutions from Gaia, combined with our bootstrap approach, provide a reliable method for studying the neighborhood demographics of HZS.

\end{document}